\documentclass[8pt]{article}  
\usepackage{hyperref}
\usepackage[all]{hypcap}
 \usepackage{doc}
\usepackage{multirow}
\usepackage{amsmath}
\usepackage[dvips]{color}
\usepackage{amsmath, amssymb, graphics}
\usepackage[table]{xcolor}
\usepackage{graphicx}
\title{{Towards 4 formalisms description of properties of \newline
the unconventional Josephson junction and unconventional SQUIDs made by putting non-superconducting strip on the top of superconducting strip}}			
\author{Krzysztof Pomorski $^{1,2}$
\\
\\
$University$ $of$ $Tokyo^{1}$, $Jagiellonian$ $University^{2}$
 }		

\date{\today}
\begin{document}
\maketitle

\begin{abstract}
{We present the theoretical approach to study the unconventional Josephson junction (uJJ) made by putting the non-superconducting strip on the top of superconducting strip. We work in the framework of the Ginzburg-Landau, Bogoliubov de Gennes and Usadel equations. We solve the non-linear partial differential equations numerically for few simple cases. We obtained the eigenenergies of the uJJ by means of combined GL and BdGe method for the simplest case.

We review the similarities and new aspects of uJJ with currently known Josephson junctions.
We predict the occurence of the physical effect, which we call the topological Meissner effect.
Basing on the obtained results and current knowledge on Josephson junctions we point the future perspectives of the research on uJJs.}
\newline
\emph{\textbf{THZ electronics, superconductor-ferromagnet system,non-linear SQUID, topological Meissner effect, relaxation method for GL equations}}
\end{abstract}
\newpage
\tableofcontents
\listoftables
\newpage
\section{Motivation and general overview}
{The unconventional Josephson junction (uJJ) can be defined as by placement of non-ferromagnetic or ferromagnetic
(non-superconducting material) on the top of superconducting strip. (as depicted on the Fig.1)
The superconductor can have order parameter of any symmetry type as s, p, d, f for the case of singlet, triplet or mixed case.

The technological process necessary to produce uJJ is quite simple, since we have to evaporate (or bring in another way) the non-superconducting material on the top of given superconductor strip. Because of the simplicity of the technological process of production of single uJJ we shall be able to produce circuits of high or medium level of integration. This has particular importance for the d-wave superconductors since some of them can be superconducting in the temperature of liquid nitrogen.
Such view was suggested by the international patent by L.Gomez and A.Maeda [0] that defines the physical structure,
which we name as uJJ. Richness of physics in such structure is considerable.

UJJ can be used possibly in the circuits for THZ electronics or superconducting qubits.
If the ferromagnetic strip is being used, it shall be possible to build the tunable Josephson junction with the
magnetization of the ferromagnetic material as the tunning parameter. Then uJJs should serve as the tunable detector of electromagnetic radiation or non-linear SQUID.
Such SQUIDs should be constructed from single or two uJJs and are expected to show highly non-linear behavior with the respect to the external magnetic field or with respect to the current flowing via the junction or both these factors.
Especially interesting behaviors of non-linear SQUID is for the case when two uJJs strongly couple via the ferromagnetic stripes. In such case for the certain regime SQUID physical structure should be preventing entrance of external magnetic field to the interior of SQUID for certain values of this field.

 Also nonlinear uJJ SQUIDs built from one or two unconventional Josephson junction should have higher sensitivity
to the external magnetic field. Particularly under certain conditions the unconventional Josephson junction shall be capable of detection fraction of quantum fluxon. Full theoretical confirmation of the occurrence of such expected phenomena requires conduction of 3 dimensional numerical simulations of Ginzburg-Landau equation for the certain configurations of the external magnetic field.
 This task is beyond the scope of this paper and will be the subject of the future work.


 In this paper we use mainly the approach based on the combined Ginzburg-Landau (GL) and Bogoliubov de Gennes (BdGe) formalisms to get the properties of the uJJ.
Such choice is due to the fact that relaxation method allows for the quick solution of the Ginzburg-Landau equation in quite many cases what was confirmed in the conducted numerical simulations.

 On another hand, the numerical solutions of the BdGe equations are more demanding, since they involve the necessity to find n eigenenergies and n complex eigenfucntions what is quite difficult especially in the 2 and 3 dimensional case.

 Other formalisms are also capable of even more detailed description of the unconventional Josephson junction.
 However the prize to be paid for the more detailed answers is the significant increase of the complexity of differential equations that have to be solved.
 Therefore application of more advanced formalisms than GL and BdGe remains the subject of future studies
 already being undertaken. 


\section{Mathematical statement of the problem}
In order to approach the complex case as the solution of the BdGe and GL equations for the uJJ we start from the simpler cases which also occurs in the system. Let us consider the case of the rectangular shape ferromagnetic or non-ferromagnetic non-superconducting bar on the top of superconductor.
As it is well known the superconductor has special type of physical properties compared to the non-superconducting (normal) state of the same material.
However the proximity effects modifies properties of superconductor (sc) and non-superconductor (nsc) material.

In general situation dealing with the ferromagnetic material on the top of superconductor
we have the ferromagnetic material to be in the one among many possible magnetized states what brings many mathematical cases to be accounted.

The simplest starting case is the whole system in the normal state.
Let us assume that the ferromagnetic material will have the similar conductivity as the superconducting material in the normal state.
Then the current flowing in the region underneath the ferromagnetic bar changes the magnetization of the ferromagnetic bar, which becomes nonuniform. The magnetization of ferromagnet can be assumed to be uniform only for the case of small thickness of ferromagnet.
The magnetic field in the ferromagnetic bar exerts the Lorentz force on the electrons and holes moving in the normal state. Electric current flowing via uJJ generates own magnetic field acting on current carriers by itself and acting on the ferromagnet by changing its magnetization. By certain magnetization ferromagnetic acts back on the moving electrons and holes. Therefore the resistance of the uJJ structure should depend nonlinearly on the magnitude of the electric current, magnetization of ferromagnetic bar in the absence of external electric current flowing via the system and on the presence of the external magnetic field.

Therefore for the given geometry of the normal and ferromagnetic material the determination of current-voltage characteristics is highly non-trivial and cases of ac and dc current or voltage shall be accounted.

Additional complication is the possible anisotropy of the resistance in the given structure, what is the case of the
d-wave superconductors and is also due to the evaporation of the non-superconducting material on the top of superconductor.

The situation is less complicated for the case of isotropic s-wave superconductor.

 D-wave superconductor is a kind of ceramic material in the normal state. Therefore if the uJJ is in the normal
state the current will tend to flow via the ferromagnetic material, which is less resistive than the superconductor in the normal state. It is expected to be for the case of ferromagnetic and non-ferromagnetic non-superconducting material on the top of superconductor.

\subsection{Case of rectangular shape superconductor of infinite height with dimensions a,b in GL approach and non-zero current flowing via the system for s-wave superconductor}

In order to approach the consideration of all cases for the uJJ with s or d wave symmetry of the order parameter, it is necessary to consider the simpler situation (limiting case), when we describe the state of superconductor with total current flowing via as I, when it is very far from the non-superconducting strip.

Therefore we study rectangular shape superconductor of infinite height and a and b
dimension in its smallest cross section, with total current flowing via the system equal as I.
We look for the distribution of the order parameter, vector potential  and current density in the structure.
The simplest way to approach this problem is to state and solve GL equations, which are only valid for the case of
temperature of uJJ close to critical temperature $T_{c}$.
Having the solutions of the GL equations, we might look for the more detailed physical properties as by solving BdGe, Usadel, Eilenberger or Gorkov equations for this system.
However those formalisms are significantly more complex and more difficult to be solved, so initially we will concentrate only on GL approach.

In the given system we use Cartesian coordinates as the most suitable for its description and set the z axis to be in the direction of the height of the rectangular shape superconductor (sc) of s-wave symmetry.
Since height is infinite the system is only invariant under shift in the z direction.
The system physical properties are therefore parameterized only by x and y coordinates.
Because of translational symmetry we expect the electric current to be flowing only in z direction.
Using London limit we have current density proportional to vector potential.
Therefore we have only z nonzero component of vector potential.
The boundaries of the rectangular sc are
\begin{center}

\begin{equation*}
A={(x_{A},y_{A})}=(x=a/2, y\in(-b/2,b/2)),
B={(x_{B},y_{B})}=(x=-a/2, y\in(-b/2,b/2))
\end{equation*}
\begin{equation*}
C={(x_{C},y_{C})}=(y=b/2, x\in(-a/2,a/2)),
D={(x_{D},y_{D})}=(y=-b/2, x\in(-a/2,a/2))
\end{equation*}

\end{center}
We have the complex scalar field describing the order parameter given as

\begin{equation*}
\psi(x,y,z)=|\psi(x,y)|e^{i\frac{2 e}{\hbar c}A_{z}(x,y)(z-z_{0})}
\end{equation*}
where $z_{0}$ is fixed value.
At the boundaries superconductor-vacuum for the superconducting rectangular suspended in vacuum we have
\begin{equation*}
\frac{d}{dx}\psi(A)=0, \frac{d}{dx}\psi(B)=0, \frac{d}{dy}\psi(C)=0, \frac{d}{dy}\psi(D)=0
\end{equation*}

Intuition tell us that the order parameter in the middle of the superconducting rectangular ($x_{0}$,$y_{0}$) has the maximum value and should be decreasing when we arrive to the boundary.

Thus we have

\begin{equation*}
\frac{d}{dx}|\psi(x,y)_{x=x_{0},y=y_{0}}|=0, \frac{d}{dy}|\psi(x,y)_{x=x_{0},y=y_{0}}|=0
\end{equation*}

GL equation can be written as the following
\begin{equation*}
\alpha \psi + \beta |\psi|^2 \psi+ \frac{1}{2m}(\bar{p}_{c}^2 \psi)=0
\end{equation*}

We assume that $\alpha$ has constant value in the given rectangular and zero outside. We have the canonical momentum operator $p_{c}$ given as $p_{x,c}=(\frac{h d}{i dx}-\frac{2e}{c}A_{x})$ and
$p_{c}^2=p_{x,c}^2+p_{y,c}^2+p_{z,c}^2$. We set gauge to be $\nabla A=0$ and obtain the electric current density in GL formalism as

\begin{equation*}
j=\nabla \times(\nabla \times A)=\nabla(\nabla A)- \nabla^2A=- \nabla^2A=\frac{\hbar}{2mi}(\psi^{\dag} \nabla \psi - \psi\nabla \psi^{\dag}  )-cA|\psi|^2=c_{3}A|\psi|^2
\end{equation*}
where $c_{3}$ is constant depending on the universal physical constants. Because of non-zero z component of A current density is $j(x,y,z)=j_{z}(x,y)$. Therefore we have the set of 2 dimensional non-linear partial differential equations for the real value functions $|\psi(x,y)|$ and $A_{z}(x,y)$, where $|\psi(x,y)|$ is always non-negative.
\begin{equation*}
-(\frac{d^2}{dx^2}+\frac{d^2}{dy^2})A_{z}(x,y)=c_{1}A_{z}(x,y)|\psi(x,y)|^2
\end{equation*}

\begin{equation*}
\psi(x,y,z)(\alpha  + \beta|\psi(x,y)|^2)- \frac{\hbar^2}{2m}( \frac{d^2}{dx^2}+\frac{d^2}{dy^2})\psi(x,y,z)+\frac{1}{2m}(\frac{\hbar}{i}\frac{d}{dz}-\frac{2e}{c}A_{z}(x,y))^2\psi(x,y,z)=0
\end{equation*}

The last equation is as
\begin{equation*}
\psi(x,y,z)(\alpha  + \beta |\psi(x,y)|^2)- \frac{\hbar^2}{2m} (\frac{d^2}{dx^2}+\frac{d^2}{dy^2})\psi(x,y,z)+c_{3}A_{z}(x,y)^2\psi(x,y,z)=0
\end{equation*}
where $c_{3}$ is constant.
We have

\begin{equation*}
-\frac{\hbar^2}{2m} (\frac{d^2}{dx^2}+\frac{d^2}{dy^2})\psi(x,y,z)=\frac{-\hbar^2}{2m}( (\frac{d^2}{dx^2}+\frac{d^2}{dy^2})|\psi(x,y)|)e^{i \frac{2e}{\hbar c}A_{z}(x,y)(z-z_{0})}
\end{equation*}
\[ -\frac{\hbar^2}{2m}|\psi(x,y)|e^{i \frac{2e}{\hbar c}A_{z}(x,y)(z-z_{0})}(\frac{2e}{\hbar c}A_{z}(x,y)(z-z_{0}))^2
((\frac{d}{dx}A_{z}(x,y))^2+(\frac{d}{dy}A_{z}(x,y))^2) +\]
\[
-\frac{\hbar^2}{2m}|\psi(x,y)|e^{i \frac{2e}{\hbar c}A_{z}(x,y)(z-z_{0})}(\frac{2e}{\hbar c}A_{z}(x,y)(z-z_{0}))
(\frac{d^2}{dx^2}A_{z}(x,y)+\frac{d^2}{dy^2}A_{z}(x,y))+
\]
\[
- \frac{\hbar^2}{2m}( (\frac{d}{dx}+\frac{d}{dy})|\psi(x,y)|)e^{i \frac{2e}{\hbar c}A_{z}(x,y)(z-z_{0})}
i \frac{2e}{\hbar c}(\frac{d}{dx}+\frac{d}{dy})A_{z}(x,y)(z-z_{0})
\]

The great complexity of equations can be simplified if we set $z=z_{0}$ so we obtain

\begin{equation*}
|\frac{\hbar^2}{2m} (\frac{d^2}{dx^2}+\frac{d^2}{dy^2})\psi(x,y,z)|=|\frac{\hbar^2}{2m} (\frac{d^2}{dx^2}+\frac{d^2}{dy^2})|\psi(x,y)||
\end{equation*}

We might attempt to solve two equations simultaneously or try to reduce two equations to the one more complicated equation.

We observe that we can write the magnitude of the order parameter only in the terms of vector potential $A_{z}(x,y)$ what is given below.
\begin{equation*}
|\psi(x,y)|=\sqrt{\frac{(-(\frac{d^2}{dx^2}+\frac{d^2}{dy^2})A_{z}(x,y))}{c_{1}A_{z}(x,y)}}
\end{equation*}

Thus we have the complicated two dimensional equation for the single real value function(the z-th component of vector potential that has certain constraints) of two coordinates as .

\begin{equation*}
\alpha \sqrt{\frac{(-(\frac{d^2}{dx^2}+\frac{d^2}{dy^2})A_{z}(x,y))}{c_{1}A_{z}(x,y)}} + \beta (\sqrt{\frac{(-(\frac{d^2}{dx^2}+\frac{d^2}{dy^2})A_{z}(x,y))}{c_{1}A_{z}(x,y)}})^3 \end{equation*}
\[ - \frac{\hbar^2}{2m}((\frac{d^2}{dx^2}+\frac{d^2}{dy^2}-(\frac{e}{c})^2A_{z}^2(x,y)))\sqrt{\frac{(-(\frac{d^2}{dx^2}+\frac{d^2}{dy^2})A_{z}(x,y))}{c_{1}A_{z}(x,y)}})
 =0  \]

To solve such highly non-linear equation analytically or numerically is uneasy task.

Additional constrain for the system is the value of the total current flowing via the system as

\begin{equation*}
\int_{-\frac{a}{2}}^{\frac{a}{2}}\int_{-\frac{b}{2}}^{\frac{b}{2}} dx dy j_{z}(x,y)=I
\end{equation*}
where I is the given value of total electric current flowing via the system.

Experimentally we can set certain value of total electric current I, which is below $I_{c}(T)$. The dependence of order parameter is accounted by the value of $\alpha(T)$. It should be underlined that in the superconducting state $\alpha$ is negative while $\beta$ is positive. In the normal state both $\alpha$ and $\beta$ are positive.

The solutions of the stated two dimensional GL equations is not straightforward. To exercise our method we will first solve the more simplified case and then solve the case currently described.
The simpler case under the consideration will be the reduction of dimensionality in the coordinates. If we consider indefinitely long superconducting cylinder of given radius R, with certain current flowing via the system that we have to describe the system by only one coordinate r.

\subsection{The solutions of the GL equation for the infinitely long superconducting cylinder of radius R with nonzero current flowing via the system}

The system has two symmetries: translational in the z direction and rotational. This simplifies the GL equations so they become dependent only on r.

Similarly as before we have the electric current flowing in the z direction so $j_{z}(r)$ electric current density component is nonzero and hence $A_{z}(r)$ is nonzero.

We expect the order parameter in the system to be of the following form

\begin{equation*}
\psi(r,z)=|\psi(r)|e^{i2(\frac{e}{c \hbar}) A_{z}(r)(z-z_{0})}
\end{equation*}
where $z_{0}$ is the any fixed value of z coordinate.

One of the additional constrains is that the total magnitude of the electric current flowing via the system is given as

\begin{equation*}
I=\int_{0}^{R} r dr j_{z}(r)
\end{equation*}

In cylindrical coordinates we have given the boundary condition as

\begin{equation*}
\frac{d}{dr}|\psi(r)|_{r=R}=0
\end{equation*}

Intuition tell us that the order parameter in the middle of the superconducting cylinder has the maximum value and should be decreasing when we arrive to the boundary.

Thus we have

\begin{equation*}
\frac{d}{dr}|\psi(r)|_{r=0}=0
\end{equation*}

We obtain GL equation as

\begin{equation*}
\alpha \psi(r,z)+ \beta \psi(r,z)|\psi(r)|^2 - \frac{\hbar^2}{2m}(\frac{1}{r}\frac{d}{dr}+(\frac{d^2}{dr^2}))\psi(r,z)+\frac{1}{2m}(\frac{\hbar}{i}\frac{d}{dz}-\frac{e}{c}A_{z}(r))^2\psi(r,z)=0
\end{equation*}
Similarly as before $\alpha$ is constant inside the cylinder and is zero outside.

This equation after taking derivatives and simplification is of the form

\begin{equation*}
\begin{split}
(\frac{1}{r}\frac{d}{dr}+(\frac{d^2}{dr^2}))\psi(r,z) & =  \psi(z)(\frac{1}{r}\frac{d}{dr}+(\frac{d^2}{dr^2}))\psi(r)
-(\frac{4e^2}{c^2}A_{z}(r)^2(z-z_{0})^2)\psi(r) \\
& +i(\psi(r,z)\frac{d}{rdr}A_{z}(r)\frac{2e}{\hbar}(z-z_{0})+ \\
&  \psi(z)\frac{d}{dr}\psi(r)\frac{2e}{\hbar}\frac{d}{dr}A_{z}(r)(z-z_{0}))
\end{split}
\end{equation*}

\begin{equation*}
\frac{1}{2m}(\frac{\hbar}{i}\frac{d}{dz}-\frac{e}{c}A_{z}(r))^2\psi(r,z)=-\frac{4e^2}{\hbar^2}\psi(r,z)A_{z}(r)^2 c_{1}
\end{equation*}

Therefore finally we obtain

\begin{equation*}
\alpha \psi(r,z)+ \beta \psi(r,z)|\psi(r)|^2 - \frac{\hbar^2}{2m}\psi(z)(\frac{1}{r}\frac{d}{dr}+(\frac{d^2}{dr^2}))\psi(r)-
\end{equation*}
\[
-(\frac{4e^2}{c^2}A_{z}(r)^2(z-z_{0})^2)\psi(r)+i(\psi(r,z)\frac{d}{rdr}A_{z}(r)\frac{2e}{\hbar}(z-z_{0})+ 
\]

\[
  \psi(z)\frac{d}{dr}\psi(r)\frac{2e}{\hbar}\frac{d}{dr}A_{z}(r)(z-z_{0}))
\]
\[
  -\frac{4e^2}{\hbar^2}\psi(r,z)A_{z}(r)^2 c_{1}=0 
\]

This equation can be separated to 2 equations on two real-valued functions. Particularly if $z=z_{0}$ we obtain the equation

\begin{equation*}
\alpha |\psi(r)|+ \beta |\psi(r)|^3 - \frac{\hbar^2}{2m}(\frac{1}{r}\frac{d}{dr}+(\frac{d^2}{dr^2}))|\psi(r)|=0
\end{equation*}

Its solutions shall be valid for any z since system has the translational symmetry in z direction.

It turns out that one of the most effective method to solve this equation is the relaxation method.

\subsection{Special case of the vortex presence in the superconducting cylinder of the infinite height and certain radius r}

We might assume the existence of the vortex exactly being placed inside the infinite-length superconductor so the translational and rotational symmetries are being preserved.
In such situation we have nonzero $A_{\psi}(r)$ and $A_{z}(r)$.

Then there occurs two minima of the magnitude of the order parameter at $r=0$ and $r=R$.
We have two non-zero currents components in the system as component $j_{\phi}(r)$ supporting the existence of vortex in the sc cylinder and component $j_{z}(r)$ responsible for the observed effective current I in the system.
\subsection{General scheme of the relaxation method in the solution of GL equations}

The presented cases of the GL equations can be solved numerically using very simple rule, as by the minimization of the free energy functional, which is the basic assumption in the derivation of the GL equations.
The final solution, which is the configuration of the fields $(\psi,A)$ (effectivly 5 scalar fields) in the given space, which can be 1, 2 or 3 dimensional with certain boundary condition fulfills the equations.
\begin{equation*}
\frac{\delta}{\delta \psi}F[\psi[x,y,z],\psi[x,y,z]^{\dag},A[x,y,z]]=0
\end{equation*}
and
\begin{equation*}
\frac{\delta}{\delta A}F[\psi[x,y,z],\psi[x,y,z]^{\dag},A[x,y,z]]=0
\end{equation*}

To approach the solution we need to make the initial guess of the field configuration on the given set that corresponds to the certain physical intuition. The initial guess should be not so far from the final solution.

Having the initial guess we perform the calculation of fields changes on the given lattice
\begin{equation*}
\frac{\delta}{\delta \psi}F[\psi[x,y,z],\psi[x,y,z]^{\dag},A[x,y,z]]=-\frac{\delta}{\delta t}\psi[x,y,z] \\
\end{equation*}


\begin{equation*}
\frac{\delta}{\delta A}F[\psi[x,y,z],\psi[x,y,z]^{\dag},A[x,y,z]]=-\frac{\delta}{\delta t}A[x,y,z]
\end{equation*}
In the simulation cases $\delta t=\Delta t$ is usually fixed and shall be not too big since it might affect the numerical stability of the algorithm and not too small since we would like to conduct the numerical GL solution in finite time with finite accuracy.
To see concrete numerical examples please refer to figures 7,8 and 3 and 4.
\subsection{Infinite lenght rectangular shape superconductor of dimensions a,b in terms of BdGe}

We assume the crystal isotropy(what implies the assumption of the lack of dependence of the effective electron and hole mass on direction) and s-wave order parameter and have the Hartree potential of the form as given below, where $V_{0}$ accounts for the energy necessary for the electron to leave the crystal.

\begin{equation*}
V(x,y)=V_{0}\delta(x-a/2)+V_{0}\delta(x+a/2)+V_{0}\delta(y-b/2)+V_{0}\delta(y+b/2)
\end{equation*}

The Hamiltonian of the free particle is then given as

\begin{equation*}
H=\frac{1}{2m}(-\hbar^2(\frac{d^2}{dx^2}+\frac{d^2}{dx^2})+(\frac{\hbar}{i}\frac{d}{dz}-\frac{2e}{c}A_{z}(x,y))^2)+V(x,y)
\end{equation*}

\begin{equation*}
u(x,y,z)_{n}=|u(x,y)_{n}|e^{i(k_{n,z} +\frac{2e}{c}A_{z}(x,y))(z-z_{0})}
\end{equation*}

\begin{equation*}
v(x,y,z)_{n}=|v(x,y)_{n}|e^{i(k_{n,z} -\frac{2e}{c}A_{z}(x,y))(z-z_{0})}
\end{equation*}
                   The normalization condition for $u_{n}(x,y)$ and $v_{n}(x,y)$ functions gives the additional constrain as
\begin{equation*}
           \int_{-a/2}^{a/2}\int_{-b/2}^{b/2}dxdy|v_{n}(x,y)|^2+|u_{n}(x,y)|^2=1
\end{equation*}

\begin{equation*}
H u_{n}(x,y,z)+ \Delta(x,y,z) v_{n}(x,y,z)=\epsilon_{n} u_{n}(x,y,z)
\end{equation*}

\begin{equation*}
-H^{\dag} v_{n}(x,y,z)+ \Delta(x,y,z)^{\dag} u_{n}(x,y,z)=\epsilon_{n} u_{n}(x,y,z)
\end{equation*}

We set the effective coupling constant to be constant inside the superconducting crystal as $V_{1}$ and 0 outside.
We set the Fermi-Dirac distribution function to be $f(\epsilon_{n})=1/(+e^{\epsilon_{n}/kT})$ so we have

\begin{equation*}
\Delta(x,y,z)=-V_{1}\sum_{n} u_{n}(x,y,z)v_{n}(x,y,z)^{\dag}(1-2f(\epsilon_{n}))=\Delta(x,y)\Delta(z)=
\end{equation*}
\[
=-V_{1}\sum_{n}|u_{n}(x,y)||v_{n}(x,y)|(1-2f(\epsilon_{n}))e^{\frac{e}{\hbar}A_{z}(x,y)(z-z_{0})}
\]

We work in the gauge so $\nabla A=0$.

We have given the additional equation for vector potential as

\begin{equation*}
\mu j_{z}(x,y)=\nabla^2 A_{z}(x,y)=\sum_{n}(|u_{n}(x,y)|^2-|v_{n}(x,y)|^2)(k_{z,n}-(2e/c)A_{z}(x,y))
\end{equation*}

The last equation can be written in terms of the matrix and eigenvalue equation. Here the eigenfunctions are A.

The last constrain to be fulfilled is that total current flowing via the system is I and given as

\begin{equation*}
\int_{-a/2}^{a/2}\int_{-b/2}^{b/2}j(x,y)dxdy=I
\end{equation*}

\subsection{Infinite length superconductor bar of a,b dimensions  in terms of Usadel equation}
The Usadel formalism is capable of dealing with the situations when the superconductor is dirty. Here we assume that
there is no current flowing via the system. We assume the following parametrization function $\Theta_{n}(x,y)$ of normal G and anomalous F propagators given as
\begin{equation*}
G(x,y)_{n}=cos(\Theta_{n}(x,y)),F(x,y)_{n}=sin(\Theta_{n}(x,y))
\end{equation*}
We can also use Riccati parametrization given as

\begin{equation*}
G_{n}(x,y)=\frac{1-a_{n}(x,y)a_{n}(x,y)^{*}}{1+a_{n}(x,y)a_{n}(x,y)^{*}}, F_{n}(x,y)=\frac{2a_{n}(x,y)}{1+a_{n}(x,y)a_{n}(x,y)^{*}}
\end{equation*}

where $a_{n}(x,y)$ is the complex function.

Both parameterizations automatically fulfills the normalization condition for normal and anomalous propagators given as
\begin{equation*}
     F_{n}(x,y)F_{n}(x,y)^{\dag}+|G_{n}(x,y)|^2=1
\end{equation*}

Here $\omega_{n}=\pi T(2n+1)$ is the Matsubara frequency and n is the integer number.

The boundary conditions are as following

\begin{equation*}
\frac{d}{dx}F_{n}(A)=0, \frac{d}{dx}F_{n}(B)=0, \frac{d}{dy}F_{n}(C)=0, \frac{d}{dy}F_{n}(D)=0
\end{equation*}

We have the Usadel equation as given below

\begin{equation*}
2\Delta(x,y)G_{n}(x,y)=\frac{D}{2}(F_{n}(x,y)\nabla^2G_{n}-G_{n}\nabla^2F_{n})+2\omega_{n}F_{n}(x,y)
\end{equation*}

This equation is further simplification of the Eilenberger formalism, which is the specific case of the Gorkov formalism. Here D is the constant that is parameterized by the relaxation time of the scattering of electrons on the impurities.

We have also the expression for the density of states given as

\begin{equation*}
N(x,E_{n})=N(0) Re (cos(\Theta_{n}(x)))
\end{equation*}
where the $N(0)$ is the DOS at the Fermi level.

If $F_{n}$ functions are small as in the case of the superconductor close to the critical temperature $T_c$ and $G_{n}$ functions are close to 1. Then simplified and linearized Usadel equation is of the form

    \begin{equation*}
                 2\Delta(x,y)=-\frac{D}{2}\nabla^2F_{n}(x,y)+2\omega_{n}F_{n}(x,y)
    \end{equation*}

The variation of $\Delta$ function compared to $F$ variation it is small. Therefore locally $\Delta$ can be considered to be constant. We introduce here the function $F_{1}(x,y)=2\omega_{n}F_{n}(x,y)-\Delta(x,y)$
Such equation is the easy to be solved. For example it can be written in the matrix form and then look for the
eigenvalues and eigenfunctions of this matrix.
We have the self-consistency relation in the way as

     \begin{equation*}
           \Delta(x,y)=\pi T  \sum_{n=-\infty}^{+\infty} F(x,y,\omega_{n})_{n}
     \end{equation*}

We need first to have the order parameter distribution as given initially by GL formalism.
Then we obtain n $F_{n}$ functions. Having those functions we obtain the new order parameter. We continue this process until the order parameter will not change in the next iteration. This means that self-consistency is obtained. In such way we obtain the corrections of the order parameter to the GL equations.

If we assume that there is non-zero current existing in the system in z direction then it modifies our parametrization and the equation of motion as by

\begin{equation*}
G(x,y)_{n}=cos(\Theta_{n}(x,y)),F(x,y,z)_{n}=sin(\Theta_{n}(x,y))e^{i\frac{2e}{\hbar c} A_{z}(x,y)(z-z_{0})}
\end{equation*}

\subsection{Simple method of solving GL, BdGe, Usadel equations on sc square}

We present the numerical method of solving nonlinear differential equation of the form

\begin{equation*}
\frac{d^2}{dx^2}+\frac{d^2}{dx^2}f(x,y)=g(f(x,y))
\end{equation*}

where $g(,)$ is a function of function f(x,y).
This method can easily be generalized to 3 dimensional case (equation of the cubid).

We assume that the superconductor is a square what imposes certain the symmetry conditions on the GL solutions.
We will use 2 dimensional grid represented by the table where the x direction is the table vertical direction
and y direction is the table horizontal direction.

We use the simple second derivative approximation and set $\Delta x= \Delta y = \Delta h$ as

\begin{equation*}
(\frac{d^2}{dx^2}+\frac{d^2}{dy^2})\psi(x_{i},y_{j})=\frac{\psi(x_{i+1},y_{j})-2\psi(x_{i},y_{j})+\psi(x_{i-1},y_{j})}{\Delta x^2}+\frac{\psi(x_{i},y_{j+1})-2\psi(x_{i},y_{j})+\psi(x_{i},y_{j-1})}{\Delta y^2}=
\end{equation*}
\[   = \frac{1}{\Delta h^2}
(-4\psi(x_{i},y_{j})+\psi(x_{i+1},y_{j})+\psi(x_{i-1},y_{j})+\psi(x_{i},y_{j+1})+\psi(x_{i},y_{j-1}))
\]
We assume that in the geometrical center of the sc square $\psi(x_{0},y_{0})=a$ and that in the neighborhood
places the GL takes values b and c as depicted below.
\begin{center}
\begin{tabular}{|c|c|c|}
  \hline
  c & b & c  \\
  \hline
  b & a & b  \\
  \hline
  c & b & c \\
\hline
\end{tabular}
\end{center}

Having the knowledge of a immediately we get the knowledge of b and c.
Extending our neighborhood of the initial center point we get the following grid of values.
\begin{center}
\begin{tabular}{|c|c|c|c|c|}
 \hline
g & e & f & e & g  \\
  \hline
e & c & b & c & e  \\
  \hline
f & b & a & b & f  \\
  \hline
e & c & b & c & e \\
\hline
g & e & f & e & g \\
\hline
\end{tabular}
\end{center}

First we establish values f. Then we compute values of e. After this we compute the value g.
Having these values computed we extend our initial grid in similar fashion and compute all the next values.

In such way we might determine all values of the GL on the square. The last thing we need to check is to confirm that
the boundary conditions at the sc-vacuum interface occurs. If this is not the case we have to take another value in the square center.

The simple algorithm is also capable of solving more complex systems having certain symmetries, for
example the superconducting strip surrounded by ferromagnetic layer or non-ferromagnetic layer as given below.

\begin{center}
\begin{tabular}{|c|c|c|c|c|}
 \hline
fe & fe & fe & fe & fe  \\
  \hline
fe & sc & sc & sc & fe  \\
  \hline
fe & sc & sc & sc & fe  \\
  \hline
fe & sc & sc & sc & fe \\
\hline
fe & fe & fe & fe & fe \\
\hline
\end{tabular}
\begin{tabular}{|c|c|c|c|c|}
 \hline
fe & fe & fe & fe & fe  \\
  \hline
fe & sc & sc & sc & fe  \\
  \hline
fe & sc & fe & sc & fe  \\
  \hline
fe & sc & sc & sc & fe \\
\hline
fe & fe & fe & fe & fe \\
\hline
\end{tabular}
\begin{tabular}{|c|c|c|c|c|}
 \hline
n & n & n & n & n  \\
  \hline
n & sc & sc & sc & n  \\
  \hline
n & sc & fe & sc & n  \\
  \hline
n & sc & sc & sc & n \\
\hline
n & fe & n & n & n \\
\hline
\end{tabular}
\end{center}

We assume that in the presented structures the magnetization of the fe is due to the magnetic field produced by the current flowing via the superconductor.
\subsection{Infinite length cylindrical shape superconductor of radius R in terms of Usadel formalism}

We assume the following parametrization function $\Theta_{n}(r)$ of normal G and anomalous F propagators given as
\begin{equation*}
G(r,\omega_{n})_{n}=cos(\Theta_{n}(r)),F(r)_{n}=sin(\Theta_{n}(r))
\end{equation*}

Such parametrization automatically fulfills the normalization condition for normal and anomalous propagators G and F.

We have the Usadel equation given as

    \begin{equation*}
                2\Delta(r)G_{n}(r)=-\frac{D}{2}(G_{n}(r)(\frac{d^2}{dr^2}+\frac{d}{rdr})F_{n}(r)-F_{n}(r)(\frac{d^2}{dr^2}+\frac{d}{rdr})G_{n}(r))+2\omega_{n}F_{n}(r)
    \end{equation*}

After using the parametrization for the case of zero electric current and zero magnetic fields in the system we have

\begin{equation*}
2\Delta(r)cos(\Theta_{n})(r)=-\frac{D}{2}(cos(\Theta_{n})(r)(\frac{d^2}{dr^2}+\frac{d}{rdr})
sin(\Theta_{n}(r))
\end{equation*}

\[
-sin(\Theta_{n}(r))(\frac{d^2}{dr^2}+\frac{d}{rdr})cos(\Theta_{n}(r))+2\omega_{n} sin(\Theta_{n}(r))
\]
Such equation is second order non-linear ordinary differential equation (ODE) and can be solved numerically quite easily. We have the boundary conditions given as
$\frac{d}{dr}F_{n,R=r}(r)=0$.
By choosing certain discrete step $\Delta r$ and by guessing the first value at the r=R we move towards the center
computing all values on the lattice. After reaching the center we compute the derivative. If it is non-zero
we start guessing from another number until this condition is satisfied.

The final solution should be of the form that $\Delta(r)$ is maximal at r=0 and in monotic way decreases as we move towards r=R.

Having evaluated $\Theta_{n}(r)$ for $r \in (0,R)$ for given n we change index n. Then we compute the new order parameter. After some iterations self-consistency is achieved.

We can also account for the electric current density flow, as for example in z direction. Then we have

\begin{equation*}
F(r,z)=|F(r)|F(z)
\end{equation*}
where F(z) is the same phase factor as for the order parameter.

The electric current density flowing via the superconductor is given by the formula

\begin{equation*}
j_{z}(r)=\pi T D N(0) 2 i e \sum_{n}
(( F_{n}(r,z)^{\dag}(\frac{\hbar}{i}\frac{d}{dz}-\frac{2e}{c}A_{z}(r)) F_{n}(r,z))
-(F_{n}(r,z)(\frac{\hbar}{i}\frac{d}{dz}+\frac{2e}{c}A_{z}(r)) F_{n}(r,z)^{\dag}))
\end{equation*}

\[
=- \pi T D \sum_{n} c_{1}^2 |F(r)_{n}|^2 A_{z}(r)
\]

where N(0) is the density of states at the Fermi level and $c_{1}$ is the constant depending on the fundamental
physical constants.

To find solution we need to deal with two coupled differential equations.

\subsection{Infinite length cylindrical shape superconductor of radius R in terms of BdGe}

We have the Hartree-Fock potential given as

\begin{equation*}
V(r)=V_{0}\delta(r-R)
\end{equation*}

We have the order parameter as given by

\begin{equation*}
\Delta(r,z)=\Delta(r)\Delta(z)=|\Delta(r)|e^{i2\frac{e}{\hbar}A_{z}(r)(z-z_{0})}
\end{equation*}

\begin{equation*}
u_{n}(r,z)=|u(r)_{n}|e^{i\frac{e}{\hbar}A_{z}(r)(z-z_{0})}
\end{equation*}

\begin{equation*}
v_{n}(r,z)=|v(r)_{n}|e^{-i\frac{e}{\hbar}A_{z}(r)(z-z_{0})}
\end{equation*}

\begin{equation*}
 H=\frac{-\hbar^2}{2m}(\frac{d^2}{dr^2}+\frac{d}{r dr})+\frac{\hbar^2}{2m}(k_{z}(r)-(e/c)A_{z}(r))^2
\end{equation*}

We have the BdGe equations in terms of
\begin{equation*}
 \frac{\hbar^2}{2m}((\frac{d^2}{dr^2}+\frac{d}{r dr})+(k_{z}(r)-(e/c)A_{z}(r))^2) u_{n}(r,z)+\Delta(r,z)v_{n}(r)=\epsilon_{n}u_{n}(r,z)
\end{equation*}

\begin{equation*}
 \frac{\hbar^2}{2m}((\frac{d^2}{dr^2}+\frac{d}{r dr})+(k_{z}(r)+(e/c)A_{z}(r))^2) v_{n}(r,z)+\Delta(r,z)^{\dag}u_{n}(r)=\epsilon_{n}v_{n}(r,z)
\end{equation*}

The self-consistency equation gives

\begin{equation*}
|\Delta(r)|=-V_{1}\sum_{n}|u_{n}(r)||v_{n}(r)|(1-2f(\epsilon_{n}))
\end{equation*}

We have the Maxwell equation correlating the rotation of the magnetic field with the current density as

\begin{equation*}
\mu j_{z}(r)=(\nabla \times \nabla \times A_{z}(r))=\nabla^2 A_{z}(r)=\mu \frac{\hbar}{m}(k_{z}-\frac{2e}{c}A_{z}(r))\sum_{n}(|u_{n}(r)|^2-|v_{n}(r)|^2)
\end{equation*}

\begin{equation*}
\frac{d}{dr}u_{n}(r,z)=(\frac{d}{dr}|u_{n}(r)|)e^{i\frac{2e}{c}A_{z}(r)(z-z_{0})}+|u_{n}(r)|)e^{i\frac{2e}{c}A_{z}(r)(z-z_{0})}i\frac{2e}{c}(z-z{0})\frac{d}{dr}A_{z}(r)
\end{equation*}

\begin{equation*}
\frac{d^2}{dr^2}u_{n}(r,z)=(\frac{d^2}{dr^2}|u_{n}(r)|)e^{i\frac{2e}{c}A_{z}(r)(z-z_{0})}+(|u_{n}(r)|)e^{i\frac{2e}{c}A_{z}(r)(z-z_{0})}(\frac{-4e^2}{c^2}A_{z}^2(r)(z-z_{0})^2)
\end{equation*}


\[ +(\frac{d}{dr}|u_{n}(r)|))e^{i\frac{2e}{c}A_{z}(r)(z-z_{0})}i\frac{2e}{c}(z-z{0})\frac{d}{dr}A_{z}(r)
+(|u_{n}(r)|))e^{i\frac{2e}{c}A_{z}(r)(z-z_{0})}
i\frac{2e}{c}(z-z_{0})\frac{d^2}{dr^2}A_{z}(r)\]
\[
-(\frac{d}{dr}A_{z}(r))^2(\frac{2e}{c}(z-z_{0}))^2(|u_{n}(r)|))e^{i\frac{2e}{c}A_{z}(r)(z-z_{0})}
  \]

To get the simplification we set $z=z_{0}$ what significantly simplifies our equations. 








\subsection{Issue of dimensionality of uJJ}

The investigation of the uJJ shall be overtaken basing on the simplification of 3 dimensional problem to 2 dimensional problem and assumption that the lenght of non-superconducting strip, which is placed on the top of infinite area  superconductor of finite thickness. In such case the relaxation algorithm was applied to solve the GL equation in the absence of magnetic field. After we have obtained the order parameter, we plug it to the BdGe equations and look for the eigenenergies of the system. It should be underlined that the reduction of the 3D dimensional structure to 1 dimensional problem seems to be not appropriate since in such case we loose many physical properties of the system.

\subsection{The simplest case of uJJ in GL picture solved by the relaxation algorithm }
We simply the problem of uJJ to two dimensions and consider the non-superconducting strip to be non-ferromagnetic. If there are no currents in the system, the GL equation is significantly simplified and $\psi(x,y)$is real and is given as.

\begin{equation*}
\alpha \psi(x,y)+\beta \psi(x,y)^3 -\frac{\hbar^2}{2m}(\frac{d^2}{dx^2}+\frac{d^2}{dy^2})\psi(x,y)=0
\end{equation*}

We parameterize the system by a,b,c,d numbers.

$\alpha=const$ for $x\in (-a,a)$, $y\in (-b,b)$, E=(-a,-b), F=(-a,b), C=(a,b), D=(a,-b), A=(-c,b), B=(c,b),$A_{1}$=(-c,b+d),$B_{1}$=(c,b+d).

We have the following boundary conditions in the system

$\frac{d}{dx}\psi(FE)=0$, $\frac{d}{dx}\psi(CD)=0$, $\frac{d}{dy}\psi(ED)=0$,$\frac{d}{dy}\psi(FA)=0$,$\frac{d}{dy}\psi(BC)=0$, $\frac{d}{dy}\psi(AB)=\frac{1}{b_{1}}\psi(AB)$,   $\frac{d}{dy}\psi(A_{1} B_{1})=0$,$\frac{d}{dx}\psi(A A_{1})=0$,$\frac{d}{dx}\psi(B B_{1})=0$
Here $b_{1}$ is real constant value that describes the behavior of order parameter at the NS interface.

For some cases we can assume that the thickness d is infinite. We can also assume for some cases that
$a \rightarrow \pm \infty $.

The Fig.13 presents the geometrical parametrization of uJJ.

\subsection{The 2D case of uJJ in BdGe picture  }

We are given the order parameter in the form as

\begin{equation*}
\Delta(x,y)=|\Delta(x,y)|e^{i \frac{2e}{\hbar c}A_{x}(x,y)(x-x_{0})+i2A_{y}(x,y)(y-y_{0})}
\end{equation*}

\begin{equation*}
u_{n}(x,y)=|u_{n}(x,y)|e^{i A_{x}(x,y)(x-x_{0})+iA_{y}(x,y)(y-y_{0})}
\end{equation*}

\begin{equation*}
v_{n}(x,y)=|u_{n}(x,y)|e^{i A_{x}(x,y)(x-x_{0})+iA_{y}(x,y)(y-y_{0})}
\end{equation*}

where $x_{0}$ and $y_{0}$ are fixed values of x and y coordinate.
We have the Hamiltonian of the form

\begin{equation*}
H=-\frac{\hbar^2}{2m}(\frac{d^2}{dx^2}+\frac{d^2}{dy^2})+V_{H}(x,y)
\end{equation*}
where H is the Hartree-Fock effective potential for electron or hole.

We define state vector as
\[ \psi=
\begin{pmatrix}
u_{n}\\
v_{n}  \\
\end{pmatrix}
\]
and we have

$\frac{d}{dx}\psi(FE)=0$, $\frac{d}{dx}\psi(CD)=0$, $\frac{d}{dy}\psi(ED)=0$,$\frac{d}{dy}\psi(FA)=0$,$\frac{d}{dy}\psi(BC)=0$,    $\frac{d}{dy}\psi(A_{1} B_{1})=0$,$\frac{d}{dx}\psi(A A_{1})=0$,$\frac{d}{dx}\psi(B B_{1})=0$

and $u_{n}'(SN^{+})-u_{n}'(SN^{+})=Q u(SN)$ and
$v_{n}'(SN^{+})-v_{n}'(SN^{+})=Q v(SN)$

where SN denotes the interface between Sc-N and Q is constant depending on Sc and N.

\section{Properties of $\pi$ Josephson junction.}

\subsection{$\pi$ Josephson junction in the Eilenberger formalism}
Using the Born approximation we might simplify Gorkov equations to the Eilenbeger equation.
We have the following normal and anormal equasilclassical Green functions
$f=f(\omega_{n},v_{f})$,$g=g(\omega_{n},v_{f})$.
We have then the impurities present in the system on which the quasiparticles are being scattered with mean relation time given  as $\tau$. Let us consider the system as depicted on the figure 19 where we assume the problem to be one dimensional. We have the following equations in Fe region :

\begin{equation*}
v_{fx}\frac{d}{dx}g=(2 \tau)^{-1}(f<f^{\dag}>-f^{\dag}<f>)
\end{equation*}
\begin{equation*}
v_{fx}\frac{d}{dx}f=-(2\omega_{n}+ih)f+(\tau)^{-1}(f<g>-g<f>)
\end{equation*}

\begin{equation*}
-v_{fx}\frac{d}{dx}f^{\dag}=-(2\omega_{n}+ih)f^{\dag}+(\tau)^{-1}(f^{\dag}<g>-g<f^{\dag}>)
\end{equation*}
Here brackets <> denonte the averaging over the Fermi surface.
\begin{equation*}
j=2\pi T N(0) \sum_{n}<(v_{f}Im(g(\omega_{n})))>
\end{equation*}
Here $v_{f}$ means the Fermi velocity,$N(0)$ is the density of states (DOS) and h is the magnetic field strength.

In Sc region we have the equations as

\begin{equation*}
v_{fx}\frac{d}{dx}g=\Delta f - f^{\dag}\Delta^{\dag}
\end{equation*}
\begin{equation*}
v_{fx}\frac{d}{dx}f=-2\omega_{n}f+2\Delta g
\end{equation*}

\begin{equation*}
-v_{fx}\frac{d}{dx}f^{\dag}=2\omega_{n}f^{\dag}-2\Delta^{\dag} g
\end{equation*}

We use the approximation as $h \tau >>1$ what makes $<f>=<f^{\dag}>=0$.

In general case the current-phase relation can be expected to be of the form
$j(\phi) = \sum_{n=1}^{+\infty} j_{n} sin (n\phi)$, where $j_{n}$ is the n-th harmonic of Josephson current and
$\phi$ is the phase difference between superconductors.

\subsection{$\pi$ Josephson junction solved by the means of the relaxation  method}

We obtain the $\pi$-Josephson junction solved by means of the relaxation algorithm. We have assumed the constant magnetization in the system in the direction z. The problem becomes one dimensional (we parameterize system by z coordinate) and is being solved by the relaxation method. Please refer to the Fig.11.


\subsection{Ferromagnetic grain on the top of superconductor and modification of the relaxation method}

The case of the rectangular shape superconductor of dimensions a,b with non-zero current flowing via it has been solved. However in our system, we deal with the ferromagnetic material on the top of superconductor.
In such case the current flowing via the superconductor magnetizes the ferromagnetic element on the top of it.
Reversely the magnetized material creates additional magnetic field that modifies the distribution of the current in the system.

\subsection{Ferromagnetic-superconductor structure and further modification of order parameter}

The situation becomes more complicated when we have to consider not only one ferromagnetic grain but entire assembly of ferromagnetic grains. Then we have to take into account not only the interaction of each grain with superconductor, but also the interaction of grains with themselves.

\subsection{Resemblance of the unconventional Josephson junction with existing systems.}

It should be pointed that very essential for the Josephson effects are the dimensions of the system and
two characteristics numbers: superconductor coherence length $\xi$ and magnetic field penetration depth $\lambda$.

If we use the s-wave superconductor those last quantities $\xi$ and $\lambda$ do not depend on the direction.
However, in the case of the d-wave superconductors they depend on the direction and we can distinguish
ab direction and c direction so we have  $\xi_{ab}$,$\xi_{c}$ and $\lambda_{ab}$,$\lambda_{c}$.

We have two main regimes when the size of the non-superconducting element is comparable with  $\xi_{ab}$ or $\xi_{c}$ and when it is much bigger. In the first case we will have the signle Josephson junction and in the second case we will have the double Josephson junction. The situation becomes even more complicated when we use the ferromagnetic material as the non-superconducting material.

When the uJJ is being placed in the perpendicular magnetic field than it shall have certain features of the  $\pi-$ Josephson junction.

However if the material is not being magnetized than it has more resemblance to a weak link and we have the system
as $Sc_{1}-Sc_{2}-Sc_{1}$.

All type of the Josephson junctions occurring in our system are field induced Josephson junctions. In some sense they belong to the new class of Josephson junctions. However they can be approximated by already known Josephson junctions. We have at least 5 limiting cases given below.

\begin{center}
\begin{tabular}{|c|c|c|c|c|c|}
  \hline
 \textbf{Case} & \textbf{Non-sc} & \textbf{Magnetiz. direct.} & \textbf{Ext. mag field} & \textbf{a/b ratio} & \textbf{Final effect} \\
    \hline
 1 &  Fe & $\bot$ to surf & $\bot$ to surf & $\approx$ 1 & ScFeSc \\
    \hline
 2 &  Fe & - & $\bot$ to surf & $\approx$ 1 & ScFeSc \\
    \hline
 3 &  Fe &  $||$ to surf  & - & $<< 1$ & $Sc_{1}Fe_{\uparrow}Sc_{2}Fe_{\downarrow}Sc_{1}$ \\
    \hline
 4 &  Non-Fe & - & - & $\approx 1$ & Sc-N-Sc \\
    \hline
 5 &  Non-Fe & - & $\perp$ to surf & $\approx 1$ & Sc-N-Sc \\
  \hline
\end{tabular}
\end{center}

It should be underlined that the uJJ shall be approximated by the networks of the Josephson junction.
The most simple approximation is about one dimensional network of the Josephson junctions; the more advanced
approximation is two or three dimensional network of Josephson junctions.

Some of the Josephson junctions in this network are being coupled.
One of the simples model representing the coupled Josephson junctions is give as below

\begin{center}
\begin{tabular}{|c|c|c|c|c|}
\hline
  vac & vac   & $\rightarrow$ fe $\rightarrow$ & vac  & vac \\
\hline
  sc & sc $\uparrow$  & sc & sc $\downarrow$  & sc \\
\hline
  vac & vac   & $\leftarrow$& vac  & vac \\

  \hline
\end{tabular}
\end{center}

\begin{center}
\begin{tabular}{|c|c|c|c|c|}
  \hline
  - & x JJ $\uparrow$  & - & x JJ $\downarrow$ & - \\
  \hline
\end{tabular}
\end{center}

It should be underlined that the fringe field coming ferromagnetic bar is responsible for creating two Josephson junctions.
Such results were considered by the first time by by W.Clinton.
It shall be underlined that higher external currents flowing in x direction via the system should uncouple two existing Josephson junctions so there will be no more polarization as depicted above.

\begin{center}
\begin{tabular}{|c|c|c|c|c|}
\hline
  vac & vac   & fe  B $\bigodot$ fe & vac  & vac \\
\hline
  sc $\rightarrow$ current flow direction sc & sc   & sc  B $\otimes$ sc & sc   &  sc $\rightarrow$ current flow direction sc  \\
\hline
  vac & vac   & vac & vac  & vac \\
\hline
\end{tabular}
\end{center}

It should be underlined that if the superconductor is the d-wave superconductor and if the c axis is in vertical
direction, than the last configuration of magnetic field should induce pinned Josephson vortices in the superconductor area underneath of the ferromagnetic strip.
\subsection{Andreev reflection in the system and Andreev bound states}

Because of defect of the order parameter underneath of the non-superconducting system we shall observe the occurrence of the Andreev reflection in uJJs. Even more complicated behavior shall occur when the non-superconducting strip is being magnetized. Then there suppose to occur the currents in the superconductor that will try to compensate the occurrence of the external magnetic field. The presence of the magnetic field should double the existing quantized level in the
$S_{1}-S_{2}-S_{1}$ system, where $S_{2}$ superconductor has smaller order parameter that $S_{1}$ singlet superconductor and has the occurrence of fractional triplet phase. Andreev reflections are well described in the framework of the BdGe theory. In order to determine the existence of the Andreev states in the system we have designed new algorithm for solving 2 dimensional BdGe after plugging the initial order parameter distribution from GL relaxation method.

In Sc(L)-N-Sc(R) system in N region there occurs the Andreev reflection, which can be described by the equation
\begin{equation*}
\int_{L}^{R} p_{electron}(E)dl - \int_{R}^{L} p_{hole}(E)dl + \phi - \beta(E)=2 \pi n
\end{equation*}
where n is the integer number, $\beta$ is the additional phase shift depending on the shape of scattering potential
$\Delta(r)$, $\phi$ is the phase difference between superconductors L and R. We can generalize this formula quite easily for 2 or 3 dimensional case. We can have the spin-flip in Andreev-bound states when the magnetic field in the N layer (Fe layer) is not uniform. Such process is schematically being described at the figure 14.

\subsection{Algorithm of solving the BdGe equation in 2 or 3 dimensions}
We map 2 dimensional order parameter distribution to the one dimensional vector. Basing on this we build 2 dimensional matrix representing discredited  2 linear BdGe equations among existing 3. Having the matrix given, we determine its eigenfuctions and eigen energies. Having the eigen energies and eigenfunctions as the one dimensional vector, we map
it into 2 dimensional space. Then by self-consistent equation we obtain the new order parameter in the system.
We repeat the whole process until the order parameter will not change in the next iteration.

Let us review of the simple 5 $\times$ 5 grid and look for the solution of the BdGe on it.

We have map two dimensional matrix
\newline

\begin{center}
\begin{tabular}{|c|c|c|c|c|}
  \hline
  $X_{5,1}$ & $X_{5,2}$ & $X_{5,3}$ & $X_{5,4}$ & $X_{5,5}$ \\
  \hline
  $X_{4,1}$ & $X_{4,2}$ & $X_{4,3}$ & $X_{4,4}$ & $X_{4,5}$ \\
  \hline
  $X_{3,1}$ & $X_{3,2}$ & $X_{3,3}$ & $X_{3,4}$ & $X_{3,5}$ \\
  \hline
  $X_{2,1}$ & $X_{2,2}$ & $X_{2,3}$ & $X_{2,4}$ & $X_{2,5}$ \\
  \hline
  $X_{1,1}$ & $X_{1,2}$ & $X_{1,3}$ & $X_{1,4}$ & $X_{1,5}$ \\
  \hline
\end{tabular}
\end{center}
into one dimensional vector as given below. Then the parametrization by two indices is changed into parametrization by one index.
\begin{table}[ht]
\caption{Two dimensional grid for $X_(x,y)$ functions that can represent $u_{n}(x,y), v_{n}(x,y)$ functions} 
\centering 
\begin{tabular}{c c c c} 
\hline\hline 
\hline 
$X_{0}$ $\rightarrow$ & $X_{1}$ $\rightarrow$ & $X_{2}$ $\rightarrow$ & $X_{3}$ $\downarrow$\\ 
$\downarrow$ $X_{7}$ $\leftarrow$ & $X_{6}$ $\leftarrow$ & $X_{5}$ $\leftarrow$ & $X_{4}$ \\ 
$X_{8}$ $\rightarrow$& $X_{9}$ $\rightarrow$ & $X_{10}$ $\rightarrow$ & $X_{11}$ $\downarrow$\\
 $X_{15}$ $\leftarrow$ & $X_{14}$ $\leftarrow$ & $X_{13}$ $\leftarrow$ & $X_{12}$ \\

\hline 
\end{tabular}
\label{table:nonlin} 
\end{table}
Mapping 2 dimensional matrix to one dimensional vector allow us to write 2 among 3 BdGe equations in the compact way.
Now we need to make one longer vector representing u and v functions as


\[ \psi=
\begin{pmatrix}
u_{n}(1) \\
\cdots  \\
u_{n}(N_{1}) \\
v_{n}(1) \\
\cdots  \\
v_{n}(N_{1}) \\
\end{pmatrix},
A \psi = \epsilon_{n} \psi
\]
We need to find now the structure of the matrix A. On its diagonal is has the order parameter mapped to the one-dimensional vector in the same fashion as u and v functions.  Now we need to represent the action of
\begin{equation*}
\frac{2m}{\hbar^2}  H u_{n,i,j} =\frac{(u_{n,i+1,j}-2u_{n,i,j}+u_{n,i-1,j})}{dx^2}+
\frac{(v_{n,i,j+1}-2v_{n,i,j}+v_{n,i,j-1})}{dy^2}
\end{equation*}
We find the structure of matrix numerically and than we find its eigevalues and eigenfunctions.

\subsection{The occurrence of the vortices in the uJJs}

We should account for the occurrence of vortices in uJJs. This is especially to be expected when we have the occurrence of ferromagnetic strip on the top of the superconductor and the presence of the external magnetic field.
The situation becomes even more complicated when we consider the d-wave superconductor with ab-plane in parallel
to the external ferromagnetic strip. Than we will have the possible occurrence of the Josephson and Abrikosov vortices and pancake vortices as well. What is interesting, if we switch external magnetic field perpendicular to the superconductor surface that we might have the switching of the Josephson to Abrikosov vortices. Since Abrikosov vortices have smaller core than Josephson vortices, for certain magnetic field magnitude, they will allow for higher superconducting current flow and thus contribute to possible increase of the critical current.

\subsection{Vortices in the superconductor-ferromagnet system}

\subsubsection{Abrikosov vortex in the superconductor}
We describe the radial defect of the order parameter caused by magnetic field punching via the superconductor.
Than we have the occurrence of non-zero vector potential as given by $A_{\phi(r)}$ and order parameter of the form
\begin{equation*}
\psi(r,\phi)=|\psi(r)|e^{i \phi r A_{\phi}(r)}
\end{equation*}.
As before we fix gauge $\nabla A=0$ so we have
$\nabla \times \nabla \times A=-\nabla^2 A(r)=-(\frac{d^2}{dr^2}+\frac{1}{r}\frac{d}{dr}) A_{\phi}(r)=\frac{e
\hbar}{2m}(|\psi(r)|^2)A_{\phi}(r)$ and $B_{z}=\frac{d}{dr}(rA_{\phi}(r))$.
We have the GL equation as for $\phi=0$,
$\alpha |\psi(r)| + \beta |\psi(r)|^3-\frac{\hbar^2}{2m}(\frac{d^2}{dr^2}+\frac{1}{r}\frac{d}{dr})|\psi(r)|+c_{3}A_{\phi}(r)^2=0$
Since the system has the rotational symmetry than the solution of the given equation is valid for all cases of $\phi$.

We can go further and describe the occurrence of the vortex for the two dimensional system as by assuming the dependence of the $\psi(r,z,\phi)=|\psi(r,z)|e^{i \phi A_{\phi}(r,z)}$

$\nabla \times \nabla \times A=-\nabla^2 A(r)=-(\frac{d^2}{dr^2}+\frac{1}{r}\frac{d}{dr}+\frac{d^2}{dz^2}) A_{\phi}(r)=\frac{e \hbar}{2m}(|\psi(r)|^2)A_{\phi}(r)$
We have
$\alpha |\psi(r,z)| + \beta |\psi(r,z)|^3-\frac{\hbar^2}{2m}(\frac{d^2}{dr^2}+\frac{1}{r}\frac{d}{dr}+\frac{d^2}{dz^2})|\psi(r,z)|+c_{3}A_{\phi}(r,z)^2=0$

Having those equations we can describe the properties of the systems as finite thickness superconductor with magnetic field perpendicular to its surface. We can also describe the vortex in the superconductor-ferromagnet system or vortex in the superconductor-normal metal system. The relaxation method can be used to solve the given set of differential equations.





One of the interesting cases to be studied is given below.
\begin{center}
\begin{tabular}{|c|c|c|c|c|c|c|}
 \hline
sc2 & sc1 & sc1 & sc1 & sc1 & sc1 & sc2 \\
  \hline
sc2 & sc1 & vortex core & sc1 & vortex core & sc1 & sc2 \\
  \hline
sc2 & sc1 & sc1 & sc1 & sc1 & sc1 & sc2 \\
  \hline
sc2 & sc1 & vortex core & sc1 & vortex core & sc1 & sc2\\
\hline
sc2 & sc1 & sc1 & sc1 & sc1 & sc1 & sc2\\
\hline
\end{tabular}
\end{center}

Similar case to the given above can be accounted by the simple algorithm described earlier.
What we need to do it to have the square symmetry of the system.

\begin{center}
\begin{tabular}{|c|c|c|c|c|c|c|}
\hline
sc2 & sc2 & sc2 & sc2 & sc2 & sc2 & sc2 \\
 \hline
sc2 & sc1 & sc1 & sc1 & sc1 & sc1 & sc2 \\
  \hline
sc2 & sc1 & vortex core & sc1 & vortex core & sc1 & sc2 \\
  \hline
sc2 & sc1 & sc1 & sc1 & sc1 & sc1 & sc2 \\
  \hline
sc2 & sc1 & vortex core & sc1 & vortex core& sc1 & sc2\\
\hline
sc2 & sc1 & sc1 & sc1 & sc1 & sc1 & sc2\\
\hline
sc2 & sc2 & sc2 & sc2 & sc2 & sc2 & sc2 \\
 \hline
\end{tabular}
\end{center}

Another interesting case is given as below

\begin{center}
\begin{tabular}{|c|c|c|c|c|c|c|}
\hline
sc2 & vortex core & vortex core & vortex core & vortex core & vortex core & sc2 \\
 \hline
sc2 & sc1 & vortex core & sc1 & vortex core & sc1 & sc2 \\
  \hline
sc2 & sc1 & vortex core & sc1 & vortex core & sc1 & sc2 \\
  \hline
sc2 & sc1 & vortex core & sc1 & vortex core & sc1 & sc2 \\
  \hline
sc2 & sc1 & vortex core & sc1 & vortex core& sc1 & sc2\\
\hline
sc2 & sc1 & vortex core & sc1 & vortex core & sc1 & sc2\\
\hline
sc2 & vortex core & vortex core & vortex core & vortex core & vortex core & sc2 \\
 \hline
\end{tabular}
\end{center}

\subsection{Tunable Josephson junction as the tool for detection of the electromagnetic radiation or for other applications}

It is possible to produce the tunable Josephson junction using the ferromagnetic strip by connecting uJJ ferromagnetic to the magnetically biasing environment. If we control externally the state of the magnetization of the ferromagnetic bar we might produce the Josephson junction whose sensitivity to the
external radiation can be tuned quite continuously. Also the tunning of the magnetization of the ferromagnetic bar on the top of superconductor changes RCSJ parameters of the circuit. Therefore we might obtain tunable Josephson junction circuits what can have its importance in future THZ electronic circuit design.

\subsection{Vortex role in the detection and emission of the electromagnetic radiation}
Basically we expect that the pinned and unpinned vortices are present in the system. Because they have the vortex core that has non-superconducting phase component therefore they can emit and absorb radiation. Vortex movement in the superconductor is the main source of the dissipation and therefore contributes to the observed electric resistance both in ac and in dc case.

In the system we might have both Abrikosov and Josephson vortices. As pointed by Franco Nori, movement of the Josephson vortices creates more dissipation than movement of the Abrikosov vortices.

Also bigger core means bigger cross-section for the absorption of the radiation and bigger cross-section for the emission of the radiation. The dynamics of vortex movement in uJJ due to the system complexity shall be rather modeled phenomenologically.

Certain simplifications of vortex movement are due to the possible reduction of dimensionality as to one or better two dimensions. This is the case of the Josephson vortices. However certain effects as switching between Josephson to Abrikosov vortices or reversely can be done only in the case of 3 dimensional considerations.

Also pancake vortices are possible to occur in the system and they are a kind of non-linear combinations of Abrikosov and Josephson vortices.

\subsection{Presence of single and triplet component of order parameter}

It shall be underline that the presence of the ferromagnetic bar shall induce the existence of triplet component
in the neighborhood of it. In the case of singlet order parameter the magnetic field is the factor which destroys
the Coopair pairs, which are the carriers of the superconductor state. Such situation is not occurring in the case
of the triplet superconductivity. It might turn out that the coming singlet Cooper pair is partly converted to
triplet Coopair pair, the triplet Coopair pair is then partly converted to singlet Coopair pair as well.
In such case, we shall observed bigger Josephson current than those expected from the s-wave $\pi$ Josephson junctions.
Please refer to the figure 12.

\subsection{The methodology to compute the Josephson current}

\subsubsection{The tunneling Josephson current}

The most known type of Josephson junction is the tunneling Josephson junction of rectangular geometry.
This is due to the fact that for the simple geometry we can describe simple tunneling model.

We have given the Hamiltonians of the left $H_{L}$ and right $H_{R}$ and superconductor as

\begin{equation*}
H_{L,R}=\sum_{n,s}\epsilon_{k,s}c^{\dag}_{k,s}c_{k,s}+\sum_{k,k',s,s'}V(k,k',s,s')
c^{\dag}_{k,s}c^{\dag}_{k',s'} c_{k',s'} c_{k,s}
+\sum_{n,k,q}c^{\dag}_{k,s}c_{k,s} \mu \sigma_{s}B_{s}(x,y,z)
\end{equation*}
where $\sigma_{i}$ are the Pauli matrices.

In general case we have the Hamiltonian H describing the Josephson junction in the form as
\begin{equation*}
H  =   H_{L} + H_{R} + H_{T}
\end{equation*}

Here the tunneling Hamiltonian is given as

\begin{equation*}
H_{T}=\sum_{k,q}(T_{lr}^{kq}c^{\dag}_{k}d_{q}+T_{rl}^{qk}d^{\dag}_{q}c_{k})
\end{equation*}

Hermicity of $H_{T}$ gives $(T_{lr}^{kq})^{\dag}=T_{rl}^{qk}$. Therefore we have

\begin{equation*}
H_{T}=\sum_{k,q}T_{lr}^{kq}c^{\dag}_{k}d_{q}+T_{lr}^{kq \dag}d^{\dag}_{q}c_{k}
\end{equation*}


where the Hamiltonian accounting for the creation(annihilation) of particle on the left side superconductor is $d^{\dag}$ ($d$) and on the right side is $c^{\dag}$ ($c$).

The $T_{lr}^{kq}$ is the tunneling matrix between left and right superconductor.

We might compute the Josephson junction properties by the computation of the partition function defined as
$Z=Tr(e^{-\frac{H}{kT}})$.

In the Hamilton formalism we have the relations between conjugate variables q and p given as

\begin{equation*}
\frac{d}{dt}q(t)=\frac{d}{dp}H[p(t),q(t),t]
\end{equation*}

\begin{equation*}
\frac{d}{dt}p(t)=-\frac{d}{dq}H[p(t),q(t),t]
\end{equation*}

We have the following relation between free energy F and Hamiltonian as given

\begin{equation*}
-\frac{1}{\beta} \frac{d}{dp}ln(Z)=<\frac{d}{dt}q>=\frac{d}{dp}F
\end{equation*}

Josephson current can be expressed in the way as
\begin{equation*}
I_{J}=-e<\frac{d}{dt}N>=\frac{d}{d \phi}F_{J}( \phi)
\end{equation*}

We have the conjugate variable given as the phase of the order parameter $\phi$ and the operator giving the numbers of the particles as N and we have $[\phi,N]=i \hbar$, where $[]$ is the anticommutator.

We can also account for the existence of the spin-current in the system.

We define $S_{k}=\sum_{q,\alpha, \beta} c_{k+q, \alpha}^{\dag} \sigma_{\alpha , \beta} c_{k, \beta}$ 
\begin{equation}
<\frac{d}{dt}S>=-\mu_{B}\frac{d}{d \Theta}F
\end{equation}

Here angle $\Theta_{n}$ is the angle in the plane perpendicular to the direction n.
Similar as before we have $[\Theta,S_{n}]=i \hbar$

The Dyson equation is given as $G^{-1}=G_{0}^{-1}+\Sigma$
and hence G normal propagator can be written as

\begin{equation}
G \approx G_{0} + G_{0}\Sigma G_{0}+ G_{0}\Sigma G_{0}\Sigma G_{0} + ...
\end{equation}

where $G_{0}$ is the normal propagator without presence of the order parameter and $\Sigma$ is the self-field, which accounts for the presence of the order parameter (self-energy) so we have

\[G_{0}=
\begin{pmatrix}
g & 0 \\
0 & -g^{*} \\
\end{pmatrix}
,\Sigma=
\begin{pmatrix}
0 & \Delta \\
\Delta^{\dag} & 0 \\
\end{pmatrix}
\]

The normal propagator structure becomes more complicated in the case of occurrence of the magnetic field as

\[g=
\begin{pmatrix}
g_{\uparrow} & 0 \\
0 & g_{\downarrow} \\
\end{pmatrix}
\].

The presence of magnetic field can cause the spin-flop during tunneling process. Therefore we shall consider the following structure of the tunneling matrix as

\[T_{lr}^{kq}=
\begin{pmatrix}
T_{lr}^{kq, \uparrow \uparrow} & T_{lr}^{kq,  \uparrow \downarrow} \\
T_{lr}^{kq, \downarrow \uparrow} & T_{lr}^{kq, \downarrow \downarrow} \\
\end{pmatrix}
\]

For Sc-I-Sc (Superconductor-Isolator-Superconductor) Josephson junctions with no impurities and small tunneling current
the tunneling matrix is the identity matrix. However for the SIS junctions with
some magnetic impurities in I region the tunneling matrix is no longer identity.
It is not diagonal in the case of S-Fe-S Josephson junctions.

It should be underline that in the proximity of the singlet superconductor
to the ferromagnetic element the triplet superconducting phase shall be induced. Therefore we account the region of the superconductor very close to ferromagnetic element and also the region of the ferromagnet very close to the superconductor.

Then the form of the self-energy shall be changed and should be as

\[
\Sigma=
\begin{pmatrix}
M & \Delta \\
\Delta^{\dag} & M \\
\end{pmatrix}
\]

Here M accounts for the magnetization of the superconductor and $\Delta$ element
 consist both singlet and triplet component. The triplet component of the superconductor can be characterized as

 $\Delta=d_{t} \circ$ $\sigma$=$i\Delta_{t} \sigma_{y}$ where $d_{t}$ stands for the vector denoting 3 components of the triplet phase. In general case we ca have the mixture of single and triplet case and express  $\Delta=d_{t} \circ \sigma + d_{s} 1_{2 \times 2}$. Here $d_{s}$ is the complex scalar field.

We can also express the tunneling Josephson junction in terms of the action as

\begin{equation*}
  S=S_{L}+S_{R}+S_{T}
\end{equation*}
  where $S_{L}, S_{L}$ describes the action for the isolated left, right superconductor
  and $S_{T}$ describes the action connected with tunneling process.


For non-interacting fields we have the inverse of normal propagator $G^{-1}_{0}$ given as

\begin{equation*}
G^{-1}_{0}=diag(-i \omega_{n}+ \epsilon_{k},-i \omega_{n}+ \epsilon_{k},-i \omega_{n}+ \epsilon_{-k},-i \omega_{n}+\epsilon_{-k})
\end{equation*}

Here $\omega_{n}$ is the Matsubara frequency and $\epsilon_{k}$ is the kinetic energy of fermion gas.

We introduce fermion coherent states $\zeta$ and performing a Hubbard-Stratonovich transformation, 
we arrive at an effective action (in Euclidean time), which reads (summation over spin indices is given by $\alpha$,$\beta$)

\begin{equation*}
S_{eff}=-0.5 \int_{0}^{\beta} d \tau \sum_{k,\sigma} (\zeta^{\dag}_{k,\sigma}(\partial_{\tau}+\epsilon_{k})\zeta_{k,\sigma}+
\zeta_{k,\sigma}(\partial_{\tau}-\epsilon_{k})\zeta^{\dag}_{k,\sigma})
+\sum_{k,q,\sigma,\beta}\Delta(\alpha,\beta)_{kq}^{\dag} \zeta_{\frac{k+q}{2},\beta} \zeta_{\frac{-k+q}{2}\alpha}
)
\end{equation*}
\[
+ \Delta(\alpha,\beta)_{kq} \zeta_{\frac{-k+q}{2}\beta}^{\dag} \zeta_{\frac{k+q}{2},\alpha}^{\dag}
-\sum_{k,k',q}\Delta(k',q)_{\alpha,\beta}^{\dag} V(k,k')^{-1} \Delta(k,q)_{\beta,\alpha}
\]


%
%
%

We have the following property

\begin{equation*}
\int D[\zeta,\zeta^{\dag}]e^{S_{eff}}=e^{\beta F_{GL}}
\end{equation*}
where $D[\zeta,\zeta^{\dag}]$ accounts for the path integral.

\subsubsection{The model $\pi$ Josephson junction in the second quantization scheme}
Since uJJ have some similarity to the $\pi$ Josephson junction we would like to briefly review its physical properties. We are given the system as $sc_{1}-fe-sc_{2}$ with thin ferromagnetic layer.
The presented model is one dimensional.
Than we have the Hamiltonian of the system given as
\begin{equation*}
H=H_{sc1}+H_{sc2}+H_{fe}+H_{t}
\end{equation*}
where the $H_{sc1}$,$H_{sc2}$ and $H_{fe}$ are the Hamiltonians of the non-interacting sc1, sc2 and fe.
The interaction is being accounted in the tunneling term $H_{t}$ that has two components $H_{t1}$ and $H_{t2}$.
We have expression for both values of (j=1,2) as

\begin{equation*}
H_{tj}=T_{\uparrow,\uparrow}\sum_{k,p,q}a_{p,\uparrow}^{\dag}c_{j,k,\uparrow}e^{(p-k+q)r_{j}}+
T_{\uparrow,\downarrow}\sum_{k,p,q}S_{q}\sigma_{\uparrow,\downarrow}a_{p,\uparrow}^{\dag}c_{j,k,\downarrow}e^{(p-k+q)r_{j}}
+h.c
\end{equation*}
Here $T_{\uparrow,\uparrow}$ or $T_{\uparrow,\downarrow}$ represents the tunneling matrix and $c_{jk,\sigma}$ is the
annihilation operator of an electron with momentum k and spin $\sigma$ in superconductor j and
$a_{p,\uparrow}^{\dag}$ is the creation operator of an electron in fe.

Let us assume that the ferromagnetic bar is magnetized in parallel to the superconductor surface. This implies
that the electrons in the Coopair pairs close to the fe surface with align in the opposite direction.

Tunneling of electron with opposite spin is more difficult than for the case of the electron with spin of the same orientation. However, if the electron with opposite to the fe magnetization spin emits magnon than it can participate in the tunneling process much more easily.

Let us denote the Fourier transform of the localized spin operator with momentum q as $S^{q}=(S_{x}^{q},S_{y}^{q},S_{z}^{q})$ in the ferromagnetic material.

In magnon-assisted tunneling process, a down-spin electron in sc1 (sc2) penetrates through the barrier into ferromagnetic part as an evanenscent wave, and changes into an up-spin electron by emitting a magnon via the exchange
interaction.

The Josephson coupling energy is calculated by the change in the thermodynamic energy
$U=-kTlnZ$ where $U=Tr(e^{-H/kT})$.

We can assume that the superconductors 1 and 2 having phases $\Theta_{1}$ and $\Theta_{2}$ are uniform and hence have
the anomalous Green functions as

\begin{equation*}
F_{\omega_{n}}=N_{S}(0)e^{i\Theta_{1}}\frac{\Delta}{\sqrt{|\Delta|^2+\omega_{n}^2}}
\end{equation*}
Using the known total coupling energy we compute the Josephson current $I=\frac{2e}{\hbar}\frac{d}{d \Theta }U=I_{c}sin(\Theta)$.

According to S.Maekawa we have

\begin{equation*}
I_{c}=(-e \frac{e}{\hbar}g_{0}(k_{b}T)^2)\sum_{n=0}^{\infty}\sum_{m=1}^{\infty} e^{-\sqrt{(2\omega_{n}+\nu_{m})/(\hbar D_{f})}d}Re(\chi_{\nu_{m}}^{+-}(d)) ((\frac{\Delta}{\sqrt{|\Delta|^2+\omega_{n}^2}}
-\frac{\Delta}{{\sqrt{|\Delta|^2+(\omega_{n}+\nu_{m})^2}}})^2 )
\end{equation*}

where $\nu_{m} = 2m \pi kT$ and
$\chi_{\nu_{m}}^{+-} (q)$ is the transverse
spin susceptibility with momentum q and frequency $\nu_{m}$ and $\nu_{m}^{+-}(d)$ is the magnon propagator.

We have
\begin{equation*}
 \nu_{m}^{+-}(d) =\sum_{q}\nu_{m}^{+-}(q)e^{iq(r_{1}-r_{2})}
\end{equation*}
is the magnon propagator traveling with a distance d and
\begin{equation*}
g_{0}=(3/2)|T_{\uparrow,\downarrow}, T_{\uparrow,\uparrow}|^2 (G_{T}/G_{K})^2 / (k_{f}^2 l d)
\end{equation*}
Here $k_{f}$ is wave-vector value at the Fermi surface, d is the thickness of ferromagnetic material and d is the mean free path for the electrons.

We have with non-spin flop tunneling
\begin{equation*}
G_{T}=(2\pi e/ \hbar)|T_{\uparrow, \uparrow}|^2 N_{s}(0)N_{f}(0)
\end{equation*}
and quantum resistance is given as
\begin{equation*}
G_{K}=\frac{e^2}{\hbar}
\end{equation*}

\subsubsection{Extension of the tunneling model to the case of the uJJ}

First of all we do not have isolator in the uJJs. Therefore we shall divide it into small squares and
consider the following Hamiltonian as

\begin{equation*}
H=H_{L}+H_{R}+H_{C}+H_{T, L\rightarrow R,R \rightarrow L}+H_{T, R \rightarrow C, C \rightarrow R}+H_{T, R \rightarrow C, C \rightarrow R}
\end{equation*}
This is how the problem is being described in one dimension. However, we deal with at least two dimensions.

Therefore we need to consider at least 9 cells as given on the picture attached below.

\begin{center}
\begin{tabular}{|c|c|c|}
  \hline
  (i-1,j+1) & (i,j+1) & (i+1,j+1)  \\
  \hline
  (i-1,j) & (i,j) & (i+1,j)  \\
  \hline
  (i-1,j-1) & (i,j-1) & (i+1,j-1)  \\
\hline
\end{tabular}
\end{center}

We need to consider the interaction of every cells with another cell.
Than we have

\begin{equation*}
H= \sum_{i,j}(H_{S}^{i,j}+H_{T}^{(i,j),(i,j+1)}+H_{T}^{(i,j),(i-1,j+1)}+H_{T}^{(i,j),(i-1,j-1)} 
\end{equation*}
\[
+H_{T}^{(i,j),(i,j-1)}+H_{T}^{(i,j),(i-1,j)}+H_{T}^{(i,j),(i+1,j)}+H_{T}^{(i,j),(i+1,j+1)}+H_{T}^{(i,j),(i+1,j-1)})
\]

We have given the formula for the
\begin{equation*}
H_{S}^{i,j}=|i,j><i,j|E_{S(i,j)}
\end{equation*}
and
\begin{equation*}
H_{T}^{(i,j),(k,l)}=(t^{(i,j),(k,l)}|i,j><k,l|+|k,l><i,j|t^{(i,j),(k,l),\dag})
\end{equation*}

Alternatively we might write the Hamiltonians in the way as
\begin{equation*}
H_{S}^{i,j}=a_{i,j}^{\dag}a_{i,j}E_{S(i,j)}
\end{equation*}
and
\begin{equation*}
H_{T}^{(i,j),(k,l)}=(t^{(i,j),(k,l)}a_{i,j}^{\dag}a_{k,l}) +a_{k,l}^{\dag}a_{i,j}t^{(i,j),(k,l),*})
\end{equation*}

{This model can be easily generalized to the 3 dimensional case. The existence of the magnetic field can be accounted
by introducing the additional index to the creator and annihilator operators and to the transmission matrix as well.

In a sense the described model accounts for the existence of the Josephson junctions in the systems that are of weak-link
type. Unlike in the tunneling Josephson junctions here the tunneling energy (tunneling matrix) has value comparably
to the self-energy.

Once the Hamiltonian is constructed that it can be represented by two dimensional matrix. The eigenvalues of this
matrix are quite easily to be found and are the eigenvalues of the sysytem.

By this method we can account for the possibility of existence of the network of the Josephson junctions in the uJJs. }
\subsection{RCSJ model} 
\subsubsection{RCSJ model for the tunneling Josephson junctions}
RCSJ stands for the resistance capacitor shunted Josephson junction.
The two-fluid model can be used to describe the properties of the tunneling Josephson junction.
Phenomenologically we recognize that the existence of two phases: superfluid electrons that can move without
dissipation and the electrons in the normal state. The presence of the last component can be represented as the
resistance and capacitance connected in parallel. The superfluid component can be represented as the perfect
Josephson junction, which was introduced by Josephson.

We have the following current-phase equation for the Josephson junction given as
\begin{equation*}
i_{c}(t)=i_{sc}(t)+i_{n}(t)=i_{0}sin(\Theta(t))+\frac{1}{R}\frac{\phi_{0}(t)}{2\pi}\frac{d}{dt}\Theta(t)+
C\frac{\phi_{0}}{2\pi}\frac{d^2}{dt^2}\Theta(t)
\end{equation*}
where $\Theta$ accounts for the phase difference of the order parameter across the Josephson junction.
\subsubsection{RCSJ model for the uJJ}

It should be underlined that there should exist the RCSJ model for the uJJ. Having such model we can extend the library of the programs serving for the complex circuit design as Spice.
In comparison with the RCSJ model for the tunneling Josephson junctions, R is nonlinear and strongly depends
on the current magnitude and presence of the external magnetic field so we have R(I,B).

The capacitance can be induced placing the nonzero perpendicular magnetic field.



\subsection{Critical temperature of the uJJ} 
The presence of the non-superconducting strip on the top of superconductor lowers the critical temperature of the
superconductor. For thin-layers of Sc-n or sc-fe there are some theoretical works describing the dependence of the
critical temperature on thickness of superconducting-nonsuperconducting layers using the Usadel formalism or Eilenberger formalism.

Let us consider the one dimensional system as the superconductor-ferromagnet system of the thickness d of the superconductor and thickness of ferromagnet of $d_{1}$.

We write the Usadel equations for the system. We have very small order parameter so we assume that
$F_{n}=sin(\Theta_{n}) \approx \Theta_{n}$ and $G_{n}=cos(\Theta_{n}) \approx 1$.

Initially I consider only Sc-N system and I assume no presence of the magnetic field.
Then we have

\begin{equation*}
2 \omega_{n}F_{n}(x)=-D(F_{n}(x)\frac{d^2}{dx^2}G_{n}(x)-G_{n}(x)\frac{d^2}{dx^2}F_{n}(x))+2\Delta(x)G_{n}(x)
\end{equation*}
in the Sc layer and

\begin{equation*}
2 \omega_{n}F_{n}(x)=-D(F_{n}(x)\frac{d^2}{dx^2}G_{n}(x)-G_{n}(x)\frac{d^2}{dx^2}F_{n}(x))
\end{equation*}

After the simplifying assumptions we obtain

\begin{equation*}
2 \omega_{n}F_{n}(x)=-D(\frac{d^2}{dx^2}F_{n}(x))+2\Delta(x)
\end{equation*}

and in the N layer we have decoupled $F_{n}$ functions as

\begin{equation*}
2 \omega_{n}F_{n}(x)=-D(\frac{d^2}{dx^2}F_{n}(x))
\end{equation*}

We have the self-consistency relation given as

\begin{equation*}
\Delta  ln \frac{T}{T_{c}}=T \sum_{n}(\frac{\Delta}{\omega_{n}}-F_{n})
\end{equation*}
Here $T_{c}$ is the critical temperature of the superconductor structure.
At outer surfaces of the bilayers of Fe (N) or Sc we have
\begin{equation*}
\frac{d}{dx}F_{n}(d)=\frac{d}{dx}F_{n}(d_{1})=0
\end{equation*}

In the interface between the superconductor and normal state occurring fox x=0 we have

\begin{equation*}
\sigma_{n}\frac{d}{dx}F_{n}(0^{-})=\frac{d}{dx}F_{s}(0^{+})\sigma_{s}
\end{equation*}
where the $\sigma_{n}$ and $\sigma_{s}$ are the conductivities of the material n and s in the normal state.

We also have another boundary condition given as
\begin{equation*}
F_{n}(0^{-})=F_{sc}(0^{-})-\xi \gamma \frac{d}{dx}F_{n}(0)
\end{equation*}

Simple algorithm as used to solve two dimensional BdGe equation can be used here to solve the Usadel equations in the limit of small order parameter size.

This algorithm is however only limited to the case of thin superconductor layer for Sc-N or Sc-Fe structure,
which implies that the order parameter present in the superconductor is small, what makes the Usadel equation to
be linear and easy to be solved. For thicker Sc layer such simple strategy of solution is no longer possible.

\subsection{uJJ in the external microwave field}

If we consider the uJJ Hamiltonians and when we have the ferromagnetic strip in the system we have modification of the magnetic field by $H(x,y,z)=\mu B_{s}(x,y,z)+B_{m}(t,x,y,z)$. The last term $B_{m}(t,x,y,z)$ is sinusoidal time dependent. Its dependence on the position is rather difficult to be determined.
Also the phase in the superconductors is time dependent
$\frac{\hbar}{2e}\frac{d}{dt}\phi(t)=V(t)$ where $V(t)$ is of sinusoidal dependence.
\subsection{SQUID built on the uJJ}

We can built the superconducting SQUID basing on the uJJ. Especially interesting effects are expected to occur when the two ferromagnetic strips starts strongly couple.

\begin{center}
\begin{tabular}{|c|c|c|c|c|c|c|}
\hline
vac & vac & vac & fe & vac & vac & vac \\
 \hline
vac & sc & sc & fe & sc & sc & vac \\
  \hline
vac & sc & vac & fe & vac & sc & vac \\
  \hline
sc & sc & vac & coupling $ \updownarrow$ & vac & sc & sc \\
  \hline
vav & sc & vac & fe & vac & sc & vac\\
\hline
vac & sc & sc & fe & sc & sc & vac\\
\hline
vac & vac & vac & \textcolor{red}{fe} & vac & vac & vac \\
 \hline
\end{tabular}
\end{center}

If the coupling is strong it may cause the occurrence of the topological Meissner effect.
The external flux of magnetic field cannot enter the interior of the SQUID since it would imply the change of the currents flow in the system.
This process is schematically described below.

 \begin{center}
\begin{tabular}{|c|c|c|c|c|c|c|}
\hline
vac $\downarrow$ vac & vac $\swarrow$  vac &  vac &  vac &  vac & vac $\downarrow$ vac & vac $\downarrow$ vac \\
 \hline
vac $\downarrow$ vac & vac & vac &  vac & vac & vac & vac $\downarrow$ vac \\
  \hline
vac $\downarrow$ vac & fe B $\rightarrow$ fe& vac & vac $\rightarrow$ vac & vac & fe B $\rightarrow$ fe & vac $\downarrow$ vac \\
  \hline
vac $\downarrow$ vac & fe B $\rightarrow$ fe& $\searrow$ & vac & $\rightarrow$ & fe B $\rightarrow$ fe & vac $\downarrow$ vac \\
  \hline
vac $\downarrow$ vac & sc I $\otimes$ sc & $\downarrow$ & vac & $\uparrow$ & sc I $\otimes$ sc & vac $\downarrow$ vac \\
\hline
vac $\downarrow$ vac & sc I $\otimes$ sc & $\swarrow$ & vac & vac & sc I $\otimes$ sc& vac $\downarrow$ vac \\
\hline
 vac $\downarrow$ vac &   vac &  vac & vac  &  vac  & vac $\leftarrow$ vac & vac $\swarrow$ vac \\
\hline
\end{tabular}
\end{center}

The similar process shall occur for the SQUID built from $\pi$ Josephson junctions.

Topological Meissner effect can be induced by the external parallel magnetic field applied to the surface of uJJ or $\pi$ SQUID.

The Meissner topological effect (MTE) occurs in 3 dimensions and is rather difficult technically to be computed for
the GL formalism. However there is also possibility of having two dimensional MTE as for two very long sc-fe strips
connected at the ends of the cross section the same as depicted on the figure above. Than investigation of such effect would involve 2 dimensions only as $A_{z}(x,y),|\psi(x,y)|$. It can be done with use of 2D GL, BdGe, Usadel or Eilenberger approach.

The elongated SQUID showing the MTE seen from the top is given below. One dimension is assumed to be very big compared to all other dimensions.

\begin{center}
\begin{tabular}{|c|c|c|c|c|c|c|}
\hline
vac & vac  & vac & vac & vac & vac  & vac  \\
  \hline
 sc & fe  & fe & fe & fe & fe  & sc  \\
  \hline
 sc & vac  & vac & vac & vac & vac & sc \\
\hline
 sc & fe  & fe & fe & fe & fe & sc  \\
\hline
 vac  & vac & vac & vac &  vac & vac  & vac \\
\hline
\end{tabular}
\end{center}

The structure given below also should exhibit MTE in the case when fe element are magnetized around the superconductor.
 \begin{center}
\begin{tabular}{|c|c|c|c|c|c|c|}
\hline
vac & vac  & vac & vac & vac & vac  & vac  \\
  \hline
 fe & fe  & fe & vac & fe & fe  & fe  \\
  \hline
 fe & sc  & fe & vac & fe & sc & fe \\
\hline
 fe & fe  & fe & vac & fe & fe & fe  \\
\hline
 vac  & vac & vac & vac &  vac & vac  & vac \\
\hline
\end{tabular}
\end{center}

  We can also quite easily simulate 2 dimensional SQUID with two uJJ Fe or uJJ Ee and Nujj or 2uJJ N or $\pi-$ JJ ,uJJ and or single uJJ so on in GL formalism.

\subsection{GL equations for uJJ SQUID}
Using certain symmetries of SQUID made of s or d wave superconductors we might simply GL equations and use the relaxation method to determine its order parameter distribution and vector potential distribution.
Let us consider 2 uJJs  made of non-ferromagnetic strips.


\subsection{Entropy and specific heat of uJJ}
 Knowing the entropy we can quite easily compute the specific heat of the uJJ. We can know entropy by solving Usadel equation.

\section{Future perspectives} 

The relaxation method seems to be the proper tool to study more complex structures as depicted on figure 9 and 10.
The obtained solution of the order parameter shall be more exact when we plug the order parameter obtained from GL  solution to the self-consistent solution of BdGe algorithm.



The suggested geometry of described uJJ shall have the potential to be used for THZ Josephson junction. The given geometry shall be theoretically investigated in the detail for s, p, d and f wave order parameter in the singlet, triple case and the mixed case. What is more, the conduction of the experiments is necessary to find the properties of the uJJs and find their correspondence to the existing theories.

We shall also use Keldysh formalism to model the behavior of uJJ in the presence of the external microwave field.
It should be underlined that since the dimensions of the uJJ available experimentally are about 1$\mu m$ they will
give the spectrum of energies too closely positioned, so its applications for the qubit are rather limited. However one could exploit the existing pinned vortices and here the possibilities seems to be more promising. In order to
achieve the real application of the uJJ to the qubit we need to have the non-superconductive strip of width comparable with superconductive coherence length.


\section*{Acknowledgement}

I would like to thank for the detailed discussions to doctor L.Gomez, professor J.Spalek,
profesor N.Hayasi, professor J.Dziarmaga, professor P.Przyslupski and professor A.Maeda
 for the insightful discussions.
We look forward for the continuation of our work that is going to provide the numerical solutions for the analysis
of more realistic experimental cases. This work is partly covered by Monbusho scholarship and partly by PhD scholarship of the author.

\section*{References}
{
[0]. L.Gomez, A.Maeda,  \emph{Josephson junction and Josephson device}, International patent, (WO/2008/010569)
\newline
[1]. A.Buzdin, \emph{Proximity effects in superconductor-ferromagnet heterostructures}, Rev. Mod. Phys. 77 (2005), 935-976
\newline
[2]. N. C. Cassol-Seewald; G. Krein, \emph{Numerical simulation of Ginzburg-Landau-Langevin equations},
Braz. J. Phys. vol.37 no.1a São Paulo Mar. 2007
\newline
[3]. A.L.Fetter, J.D.Walecka, \emph{Quantum Theory of Many-Particle Systems}
\newline
[4]. Contribution of the microscopic theory of the Josephson effect in superonducting bridges.
\newline
[5]. T.W.Clinton, M.Johnson, \emph{Magnetoquenched superconducting value with bilayer ferromagnetic film for uniaxial switching},  Applied Physics Letters (2000), Vol 76, Nr 15.
\newline
[6].  K.Usadel, \emph{Generalized diffusion equation for superconducting alloys},  Physical Review Letters (1970)
\newline
[7]. K.Pomorski, \emph{Ideas for tunable detector array based on novel HTS weak links}, Poster presentation, Quantum metrology conference, Poznan (2007)
\newline
[8]. K.Pomorski, \emph{The prediction of the new superconducting phenomena in the superconductor and superconducting mesoscopic structures} , Master of Science thesis at the University of Lodz (2007),in Polish
\newline
[9]. K.Pomorski, \emph{Superconducting qudit as generalization of the superconducting qubit architecture},  (Patent in preparation).
\newline
[10]. M.Thinkham, \emph{Introduction to superconductivity},
\newline
[11]. A.Martin, \emph{Self-Consistent Transport Properties of Superconducting Nanostructures}, Czechoslovak Journal of Physics, Vol.46 (1996), Suppl. S4
\newline
[12] M.Krawiec, \emph{Current carrying Andreev bound states in a Superconductor-Ferromagnet proximity system},arXiv:/cond-mat/0302162
\newline
[13]. Liao Yan-Hua, \emph{Zeeman effects on Josephson current in d-wave superconductor/d-wave superconductor Josephson junctions}, IOP Science 1674-1056/17/5/058 (2008),
\newline
[14]. A.Buzdin, \emph{Domain wall superconductivity in hybrid superconductors-ferromagnetic structures},  arXiv: cond-mat/0305520v1
\newline
[15]. Ya.V.Fominov, \emph{Critical temperature of the superconductor/ferromagnet bilayers},  arXiv:cond-mat/0106185v2 (2001)
\newline
[16]. Excerpt from the Proceedings of the 2008 Nordic-COMSOL conference
\newline
[17].  Ch.Ishii, \emph{Josephson currents through Junctions with Normal Metal Barriers}, Progress of theoretical Physics, Vol.44, No.6. December 1970
\newline
[18]. S.Kawabata, Theory of Macroscopic Quantum Tunneling and Dissipation in High-$T_{c}$ Josephson junctions
\newline
[19]. A.Nakayma, \emph{Wave functions of Andreev bound states in superconductor/normal metal/superconductor junctions}. J. Appl. Phys. 91, 7119 (2002)
\newline
[20]. A.V.Dmitrev, \emph{On the details of the thermodynamical derivation of the Ginzburg-Landau equations},  ArXiv:cond-mat/0312094v1
\newline
[21]. S.V.Yampolskii, \emph{Magnetic dipole-vortex interaction in a bilayered superconductor/soft-magnet heterostructure},(2006) Europhys. Lett. 74 334-340
\newline
[22]. Y.Kim, \emph{Pairing in the Bogoliubov-de Gennes equations},  (Purdue University)
\newline
[23]. T.Yokohama, A.Maeda, \emph{Quantative model for the $I_{c}R$ product in d-wave Josephson junctions},  PRB B 76 (2007)
\newline
[24]. Y.Tanuma, New directions of superconducting nanostructures 2009 (NDSN 2009), 4-5 September, Nagoya Univ.
\newline
[25]. M.Kalenkov, A.Zaikin, \emph{Josephson current in balistic heterosctructures with spin active interfaces},
 arXiv:0811.0685v2
\newline
[26]. A.Zyuzin, B.Spivak, \emph{A theory of $\pi/2$ superconducting Josephson junctions},  arXiv :cond-mat/9910048v1
\newline
[27]. W.Clinton, \emph{Approximate solutions for the Bogoliubov -de Gennes equations:
Superconductor-normal metal-superconductor junctions and the vortex problem}, Physical Review B (1992)
\newline
[28]. Z.D.Genchev, T.L.Boyadjiev, \emph{On the solution of the modified Ginzburg-Landau type equation for a one-dimensional superconductor in the presence of normal layer}, Euro. Jnl of Applied Mathematics (2003), vol.14, pp.247-256.
\newline
[29]. T.Kato, A.Golubov, Y.Nakamura, \emph{Decoherence in a superconducting flux qubit with a $\pi$-junction}.  (2007)
\newline
[30]. B.J.Powell, \emph{The gap equations for spin singlet and triplet ferromagnetic superconductors},  IOP
2003 J.Phys. A.Math/ Gen. 36 9289
\newline
[31]. M.Houzet, \emph{Long range triplet Josephson effect through a ferromagnetic trilayer},  Physical Review B 76 (2007)
\newline
[32].  N.Stefankis, \emph{Charge current in a ferromagnet-triplet superconductor junctions}, IOP (2001), J.Phys. Condens.Matter 13 3643
\newline
[33]. J.Linder, \emph{Proximity effect in ferromagnet/superconductor hybrids:From diffusive to ballistic motion}, Physical Review B 79 (2009)
\newline
[34].  Y.Asano, \emph{Josephson Effect due to Odd-frequency Pairs in Diffusive Half Metals}, arXiv:cond-mat/0609566v1
(2006)
\newline
[35]. J.Kolacek, \emph{Surface charge of a flat superconducting slab in the Meissner state}, arXov:cond-mat/0111243v1
\newline
[36].E. K. Dahl1 and A. Sudbo, \emph{Derivation of the Ginzburg-Landau equations of a ferromagnetic p-wave superconductor}, arXiv:cond-mat/0609334v1 [cond-mat.supr-con] (2006)
\newline
[37]. S.Maekawa, \emph{Spin current in superconductors}, Progress of theoretical Physics Supplement No.176, 2008
\newline
[38]. B.D. Josephson, \emph{Possible new effects in superconductive tunnelling}, Phys. Lett. 1 (1962) 251-253.
\newline
[39]. R.J.Troy, \emph{A self-consistent Microscopic Theory of Superconductivity}, arXiv:cond-mat/9411099v1 (1994)
\newline
[40]. M.Houzet, \emph{Ferromagnetic Josephson Junction with Precessing Magnetization}, PRL 101,057009 (2008)
\newline
[41]. Y.Asano, \emph{Josephson current through superconductor/diffusive-normal-metal/superconductor junctions:
Interference effects governed by pairing symmetry}, Physical Review B 74, 064507 (2006)
\newline
[42]. A.Buzdin, \emph{Superconductor-Ferromagnet structures}
\newline
[43]. F.Konsechelle, \emph{Nonsinusoidal curremt-phase relation in strongly ferromagnetic and moderately disordered SFS junctions}, arXiv:0807.2560v1  cond-mat.mes-hall, (2008)
\newline
[44].M.Houzet, \emph{Domain wall superconductivity in superconductor/ferromagnetic bilayers, Nanoscale and superconductivity}, Argonne, November 14-18 2005
\newline
[45].H.Suematsu, \emph{Finite element method for Bogoliubov-de-Gennes equation: application to nano-structure superconductor}, Physica C 412-414 (2004)
\newline
[46]. A.M.Zagoskin, \emph{Mesoscopic d-Wave Qubits: Can High-$T_{c}$ Cuprates Play Role in Quantum Computation}, arXiv:cond-mat/0506039v1 (2005)
\newline
[47].M.H.S.Amin, \emph{Quasiclassical Theory of Spontaenous Currents at Surfaces and Interfaces of d-Wave Superconductors},
arXiv: cond-mat/0105486v3 (2002)
\newline
[48].J.P.Morten, \emph{Coherent and Correlsted Spin Transport in Nanoscale Superconductors}, PhD Thesis thesis (2008),
Norwegian University of Science and Technology
\newline
[49]. J.Linder, \emph{Proximity effect in ferromagnet/superconductor hybrids: From diffusive to ballistic motion},
PRB 79, 064514 (2009)
\newline
[50].D.Manske, \emph{Electronic properties of strongly correlated oxides in reduced dimensions} (theory), presentation at
ETH
\newline
[51]. G.Eilenberger, G\emph{eneral approximation method for the Free Energy Functional of Superconducting Alloys},
Zeitchrift fur Physik 190, 142-160 (1966)
\newline
[52]. D.Huertas-Hernando, \emph{Generalized boundary conditions for the circuits theory of spin-transport},
arXiv:cond-mat/0204116v1 (2002)
\newline
[53]. Golubov, \emph{The current-phase relation in Josephson junctions}, Review of Modern Physics Vol.76,No.2. (2004)
\newline
[54].JETP Letters Vol.61, No.5, (1995)
\newline
[55]. Alex Kamenev and Alex Levchenko, Keldysh technique and non–linear $\sigma$  –model:
basic principles and applications,
\newline
$http://arxiv.org/PS_cache/arxiv/pdf/0901/0901.3586v3.pdf$
\newline
[56]. L. V. Keldysh: JETP 20 (1965) 1018.
\newline
[57]. Z.G.Ivanov,\emph{Boundary conditions for the Usadel equations and properties of dirty S-N-S sandwiches},
Journal de Physique (1978), http://hal.archives-ouvertes.fr/docs/00/21/76/74/PDF/ajp-jphyscol197839C6250.pdf
\newline
[58]. $http://boulder.research.yale.edu/Boulder-2000/lectures/martinis/usadel.pdf$
\newline
[59]. Eilenberger, \emph{Transformation of Gorkov's Equation for Type II Superconductors into Transport-Like Equations},Z. Phys. 214 (1968) 195
\newline
[60]. De Gennes, \emph{Superconductivity of Metals and Alloys}, (W.A. Benjamin, New York) 1966
\newline
[61]. K. Likharev, \emph{RSFQ logic/memory family: A new josephson-junction technology for sub-terahert5clock- frequency digital systems},
\newline
$http://ieeexplore.ieee.org/iel4/77/9206/00413117.pdf?arnumber=413117$
\newline
[62]. M.H.S.Amin, \emph{Quasiclassical calculation of spontaneous current in restricted geometries},
arXviv:cond-mat/0207617v2
\newline
[63]. A.Maeda ,\emph{Nanoscale Friction: Kinetic Friction of Magnetic Flux Quanta and Charge Density Waves},
PRL 94, 077001 (2005)
\newline
[64]. F.Nori, \emph{Using Josephson Vortex Lattices to Control Terahertz Radiation:
Tunable Transparency and Terahertz Photonic Crystals}, PRL 94, 157004 (2005)
\newline
[65].H.L.Yu, \emph{Proximity effects in ferromagnet/superconductor bilayers}, Physics Letters A 339 (2005)
\newline
[66]. D.O.Gunter, \emph{Computational and experimental studies of vortex dynamics in type II superconductors}, PhD thesis at The University of Wisconsin—Milwaukee (1999)
\newline
[67].M.V.Milosevic, \emph{Vortex pinning in a superconductor film due to in-plane magnetized ferromagnets of different shapes: The London approximation}, PRB 69, 014522 (2004)
\newline
[68].Heikkila, Phys. Rev B 66, 184513 (2002)
\newline
[69]. A.P.Jauho, \emph{Introduction to the Keldysh nonequilibrium Green function technique}
\newline
[70]. Francesco Giazotto,\emph{Thermal properties in mesoscopics: physics and applications from
thermometry to refrigeration}, $http://arxiv.org/PS_cache/cond-mat/pdf/0508/0508093v4.pdf$
\newline
[71].T.Yokohama, Meissner effect in diffusive normal metal/superconductor junctions, Physica C 426-431 (2005)
\newline
[72]. L.G.Aslamazov, Josephson effect in wide supercoducting bridges, Sov.Phys. JETP Vol.41, No. 2
\newline
[73].Y.Takane, \emph{Charge Imbalance Transport in a Superconducting Wire with Spatial Variation of Pair Potential Amplitude}, J. Phys. Soc. Jpn., Vol. 78, No. 6 (2009)
\newline
[74].A.Buzdin, \emph{Direct coupling between magnetism and superconducting current in Josephson $\pi$ junction},
\newline
$http://arxiv.org/PS_cache/arxiv/pdf/0808/0808.0299v1.pdf$
\newline
[75].W.Belzig,\emph{Quasiclassical Green's function approach to
mesoscopic superconductivity} ,Superlattices and Microstructures, Vol. 77, No. 7, 1999
\newline
[76].F.Konschelle, Non-sinusoidal current=phase relations in strongly ferromagnetic and moderatly disordered SFS junctions
\newline
\newline
\newline
Fig.1.The layout of the structure is given below(picture produced by dr.L.Gomez)
\newline
Fig.2. The scheme of the computed structure of solutions of GL equation for s-wave order parameter for 2 dimensional case with the use of relaxation method. No currents and no magnetic fields are present in the system.
\newline
Fig.3. The cross section of the order parameter for the rectangular shape superconductor of infinite length (upper picture). The relaxation method used for solution GL equation gives the free energy dependence with iterations as given below.
\newline
Fig.4. The dependence of the total electric current flowing via rectangular shape superconductor of infinite height on the iteration step in the GL relaxation method. Please refer to the figure 3.
\newline
Fig.5. Scheme of 3 basic configurations of magnetic field which are due to the certain magnetization of ferromagnetic bar, which is being put on the top of superconductor
\newline
Fig.6. The uJJ magnetic field triangle, which accounts for all possible situation of magnetization of the ferromagnetic bar, with given fixed magnitude of magnetic field in geometric center of the ferromagnetic bar. One direction of this magnetic field corresponds to one point in this triangle. Changing the direction of magnetic field, but keeping its magnitude constant, we move from one to another point in this triangle. Any point of this triangle is the reference point to establish other parameters of unconventional Josephson junction (as resistance of uJJ in the limits of small currents or capacitance of the uJJ in the limit of small electric current flowing via junction).  Points A, B, C correspond to the configurations of magnetic field as given on the sub-pictures of Figure 5.
\newline
Fig.7. The stages of solution after iterations  t0,2t0, 3t0, 4t0 of the GL equation for rectangular of infinite height and finite a=b dimensions.(please see below). No currents and no magnetic fields are assumed to occur in the system.  
\newline
Fig.8. (Picture below )Distribution of the order parameter for 3D SNS system obtained by the solution of GL equation with use of relaxation algorithm. No currents and no magnetic field is assumed to occur in the system. Horizontal axes correspond to x,y coordinates and vertical axes refers to the magnitude of order parameter. z=const for the given picture.
\newline
Fig.9. The SQUIDs built with use of uJJ. The picture above describes the system showing topological Meissner effect as predicted in the Master of Science thesis of Krzysztof Pomorski (2007) at the University of Lodz (Predictions of physical phenomena in the mesoscopic superconducting structures and superconductors ). In order to prevent the occurrence of the topological Meissner effect one has to decouple the ferromagnetic strips what is depicted on the picture below.
\newline
Fig.10. The possible configuration of the magnetic field in the sc-fe system, which exhibits the occurrence of exotic vortex in the superconductor. The studies of this structure shall be in the potential of GL relaxation method. Such vortices are expected to occur in uJJ system.
\newline
Fig.11. Order parameter for the superconductor-ferromagnet-superconductor system of fixed thickness of ferromagnetic material for different values of magnetic field. We assume the magnetization of the iron to be perpendicular to the surface of superconductor. No electric current perpendicular to the superconductor surface occurs in the system.  We can observe the oscillations of the order parameter in the system as given by Buzdin . This is example of the 1 dimensional GL problem solved by relaxation method.
\newline
Fig.12. The behaviour of the Coopair pairs in the proximity of the magnetized ferromagnetic bar. One should expect the occurrence of the induced triplet phase of superconductor in the proximity of the magnetized ferromagnet. Such situation shall occur in the uJJ as well.
\newline
Fig.13. Parameterization of the unconventional Josephson junction
\newline
Fig.14. Possible Spin-flip in uJJ junction.
\newline
Fig.15. The contour plot of real valued order parameter distribution (no magnetic field and currents occur in the system) in the rectangular of infinite height and dimensions of a,b with the rectangular hole inside also infinite height and dimensions c,d. The order parameter is obtained from the relaxation method in GL formalism. Such system is preludium to study more complex superconductor geometries,  including 2 and 3 dimensional SQUID built with unconventional Josephson junctions.
\newline
Fig.16. Solution of GL equation for 3 dimensional uJJ with following parameters:Nx=Ny=Nz=20 (nr of grids elements in 3 dimensions), Lx=Ly=Lz=4 (coherence length=0.5),ybegin=5, yend=15, beginning and ends of normal strip on the left side of  picture),the number of iterations is 1000.
\newline
Fig.17. Solution of GL equation for 3 dimensional uJJ with following parameters:Nx=Ny=Nz=20 (nr of grids elements in 3 dimensions), Lx=1, Ly=Lz=4 (coherence length=0.5),ybegin=5, yend=15, beginning and ends of normal strip on the left side of  picture),the number of iterations is 1000.
\newline
Fig.18. The eigenspectrum of 2 dimensional solution of BdGe equations for uJJ with no currents and fields. Parameters from Fig.16 are being taken.
\newline
Fig.19. The scheme of pi-Josephson junction with long Fe channel. Consequently one dimensional Eilenberger equation can be solved and it gives non-sinusoidal current phase relation.
\newline
Fig.20. Multiple BdGe wavepacket reflections in SNS system. The red arrows stands for hole wavepacket, while black arrow stands for the electron wavepacket.
\newline
Fig.21. Non-uniform BdGe wavepacket reflection in uJJ.


\end{document}